%% file: ms.tex
\documentclass[twocolumn]{aastex631}

\definecolor{fuchsia}{rgb}{0.54, 0.17, 0.89}
\definecolor{azure}{rgb}{0.0, 0.5, 1.0}
\definecolor{pgreen}{rgb}{0.12, 0.3, 0.17}
\definecolor{alizarin}{rgb}{0.82, 0.1, 0.26}

\newcommand{\kms}{{\rm km~s^{-1}}}

\newcommand{\oiii}{[\textrm{O}~\textsc{iii}]}

\newcommand{\nii}{[\textrm{N}~\textsc{ii}]}

\newcommand{\simgt}{\,\rlap{\lower 3.5 pt \hbox{$\mathchar \sim$}} \raise
1pt \hbox {$>$}\,}
\newcommand{\simlt}{\,\rlap{\lower 3.5 pt \hbox{$\mathchar \sim$}} \raise
1pt \hbox {$<$}\,}
\newcommand{\Msun}{M_{\odot}}

\newcommand{\logm}{\log M_*/\Msun}

\newcommand{\ha}{${\rm H\alpha}$}
\newcommand{\hb}{${\rm H\beta}$}
\newcommand{\hg}{${\rm H\gamma}$}
\newcommand{\hd}{${\rm H\delta}$}

\newcommand{\lya}{${\rm Ly\alpha}$}

\newcommand{\Z}{{\rm [Z/H]}}

\newcommand{\ewh}{EW$_0 ({\rm H\alpha}\!+\![\textrm{N}~\textsc{ii}]) $}
\newcommand{\De}{$D_e{4000}$}

\newcommand{\ntot}{42} %
\newcommand{\ntotza}{17} %
\newcommand{\ntotzb}{25} %
\newcommand{\ntotfld}{29} %

\usepackage{amsmath}

\accepted{February 4, 2025}

\submitjournal{ApJ}

\shorttitle{Accelerated Emergence of Evolved Galaxies in Early Overdensities at $z\sim5.7$}
\shortauthors{Morishita et al.}

\graphicspath{{./}{figures/}}

\begin{document}

\title{
Accelerated Emergence of Evolved Galaxies in Early Overdensities at $z\sim5.7$
}

\correspondingauthor{Takahiro Morishita}
\email{takahiro@ipac.caltech.edu}

\author[0000-0002-8512-1404]{Takahiro Morishita}
\affiliation{IPAC, California Institute of Technology, MC 314-6, 1200 E. California Boulevard, Pasadena, CA 91125, USA}

\author[0009-0002-8965-1303]{Zhaoran Liu}
\affiliation{Astronomical Institute, Graduate School of Science, Tohoku University, 6–3 Aoba, Sendai 980-8578, Japan}

\author[0000-0001-9935-6047]{Massimo Stiavelli}
\affiliation{Space Telescope Science Institute, 3700 San Martin Drive, Baltimore, MD 21218, USA}
\affiliation{Dept. of Physics \& Astronomy, Johns Hopkins University, Baltimore, MD 21218, USA}
\affiliation{Dept. of Astronomy, University of Maryland, College Park, MD 20742, USA}

\author[0000-0002-8460-0390]{Tommaso Treu}
\affiliation{Department of Physics and Astronomy, University of California, Los Angeles, 430 Portola Plaza, Los Angeles, CA 90095, USA}

\author[0000-0001-9391-305X]{Michele Trenti}
\affiliation{School of Physics, University of Melbourne, Parkville 3010, VIC, Australia}
\affiliation{ARC Centre of Excellence for All Sky Astrophysics in 3 Dimensions (ASTRO 3D), Australia}

\author[0000-0003-3691-937X]{Nima Chartab}
\affiliation{IPAC, California Institute of Technology, MC 314-6, 1200 E. California Boulevard, Pasadena, CA 91125, USA}

\author[0000-0002-4140-1367]{Guido Roberts-Borsani}
\affiliation{Department of Physics and Astronomy, University of California, Los Angeles, 430 Portola Plaza, Los Angeles, CA 90095, USA}
\affiliation{Department of Astronomy, University of Geneva, Chemin Pegasi 51, 1290 Versoix, Switzerland}

\author[0000-0003-0980-1499]{Benedetta Vulcani}
\affiliation{INAF Osservatorio Astronomico di Padova, vicolo dell'Osservatorio 5, 35122 Padova, Italy}

\author[0000-0003-1383-9414]{Pietro~Bergamini}
\affiliation{Dipartimento di Fisica, Università degli Studi di Milano, Via Celoria 16, I-20133 Milano, Italy}
\affiliation{INAF - OAS, Osservatorio di Astrofisica e Scienza dello Spazio di Bologna, via Gobetti 93/3, I-40129 Bologna, Italy}

\author[0000-0001-9875-8263]{Marco Castellano}
\affiliation{INAF Osservatorio Astronomico di Roma, Via Frascati 33, 00078 Monteporzio Catone, Rome, Italy}

\author[0000-0002-5926-7143]{Claudio Grillo}
\affiliation{Dipartimento di Fisica, Università degli Studi di Milano, Via Celoria 16, I-20133 Milano, Italy}
\affiliation{INAF - IASF Milano, via A. Corti 12, I-20133 Milano, Italy}




\begin{abstract}
We report the identification of two galaxy overdensities at $z\sim5.7$ in the sightline of the galaxy cluster Abell~2744. These overdensities consist of \ntotzb\ and \ntotza\ member galaxies, spectroscopically confirmed with JWST NIRSpec/MSA and NIRCam/WFSS. Each overdensity has a total stellar mass of $\sim2\times10^{10}\,M_\odot$ and a star formation rate of $\sim200\,M_\odot$/yr within a central region of radius $R=2$\,Mpc (physical). 
The sensitive PRISM spectra allow us to identify six galaxies that show weak \ha+\nii\ emissions within the overdensities ($25\pm7\,\%$), whereas the fraction of such galaxies is found significantly lower ($6\pm2\,\%$) in field samples of the equivalent redshift range.
These weak emission line galaxies, dubbed as wELGs, exhibit a strong continuum break at $4000$\,\AA\ rest-frame, a characteristic feature of evolved stellar populations. The high observed fraction of wELGs in the two overdensities is consistent with the idea that high-density environments are an ideal site where galaxies can accelerate their evolutionary pace compared to field analogs. Our study pinpoints an early onset of environmental effects, already important within one billion years after the Big Bang, and provides a complementary perspective on the emergence of quenched, massive galaxies at lower redshifts. Potential contributions from black hole accretion feedback to the reduction of star formation activity are discussed, but the connection to the local environments remains unclear.
\end{abstract}

\keywords{}


\section{Introduction} \label{sec:intro}

Early galaxy overdensities mark unique spots in the universe and play critical roles in hierarchical structural formation. Starting in initial matter density peaks, by bringing together dark matter and gas flows, they allow galaxies to efficiently form and evolve. Over time, interactions among surrounding galaxies and gaseous media further enhance star formation and chemical enrichment, leading to the maturation of galaxies within these gravitationally bounded systems. Such unique environments culminate in the distinctive features of cluster galaxy populations seen in the local universe: clusters and groups are occupied by quiescent, dispersion-dominated, massive galaxies, whereas the fraction of younger, fast-rotating, actively star-forming galaxies inversely correlates with the local density \citep[e.g.,][]{dressler80,thomas05,cappellari11}.

When, where, and how such distinct features began to appear has been a central question. Observational studies have extended the redshift frontier beyond the local universe, successfully capturing the transition of cluster galaxy properties over a wide range of cosmic time, indicative of environmental processes at work \citep[e.g.,][]{dressler97,treu03,steidel05,kodama07,newman15,wang16}. The frontier now extends to $z>3$, where a substantial number of (proto-) clusters and overdensities have been identified from large surveys \citep[e.g.,][]{steidel00,capak11,casey16,lemaux18,higuchi19,toshikawa24}, and even reaches the first billion years of cosmic history {\citep{castellano18,castellano23,harikane19,hu21,laporte22,morishita23b,sun23,Herard-Demanche23,brinch23,kashino23,wang24,arribas24,helton24,eilers24,li24}.}

Recent studies found massive galaxies that reside in overdensities out to $z\sim5$ \citep{tanaka24,kakimoto24,degraaff24,jin24,alberts23}.  Intriguingly, many of these massive galaxies show little or no ongoing star formation, suggesting that the overdensities act as accelerated environments for galaxy evolution. Even at higher redshifts, a study has shown that JADES-GS-z7-01-QU, a galaxy at $z=7.3$ \citep{looser23}, is located in an excess of strong line emitters \citep[][also Daikuhara, in prep.]{helton23}. 

However, the presence of quiescent galaxy populations in early overdensities is puzzling. It even seems to contradict the expectations of large gas reservoirs \citep[e.g.,][]{cai17,damato20,umehata21} and with the fact that galaxies generally experience elevated levels of star formation therein \citep[e.g.,][]{elbaz07,cooper08,Hayashi16, shimakawa18,perez-martinez24,laishram24}. 

Furthermore, well-established environmental quenching \citep[e.g.,][]{peng10,wetzel12}, where galaxies are quenched through physical mechanisms such as ram-pressure/tidal stripping and interactions after infall, is not expected to be the dominant channel for star formation quenching at such early times for massive galaxies {\citep[e.g.,][]{dannerbauer17,lovell23}. On the other hand, \citet{alberts23} found tentative evidence for environmental quenching of low-mass ($\logm < 9.5$) galaxies.} If the ubiquity of quiescent galaxies in early overdensities is indeed confirmed, it suggests the need for quenching processes that preferentially operate in such environments. Otherwise, the origin may be attributed to self-driven processes, such as feedback from supernovae and active galactic nuclei (AGN). The current sample size of high-$z$ overdensities is small, preventing us from reaching robust conclusions.

In this paper, we report the discovery of two galaxy overdensities at $z\sim5.7$ in the sightline of the galaxy cluster Abell~2744. Spectroscopic data from multiple JWST programs identify \ntot\ member galaxies within these overdensities. The sensitive spectrographs onboard JWST enable us to measure their star formation activity through rest-frame optical emission lines and to infer their formation history from the underlying stellar populations. Remarkably, we observe a significant excess of galaxies with {weak emission lines} in both overdensities -- compared to the surrounding field -- suggesting that galaxy evolution is already influenced by these environments. By exploiting these systems as a laboratory, our study aims to provide a unique perspective on initial assembly and growth of galaxies in early overdensities, and the quenching processes therein. 

This paper is structured as follows: In Sec.~\ref{sec:data}, we present the dataset used and the details of reduction. In Sec.~\ref{sec:member}, we report the discovery of the two overdensities and their large-scale characterization. In Sec.~\ref{sec:phys}, we investigate physical properties of a subset of the spectroscopic samples and report the population of weak emission line galaxies, which carries a critical role in this paper. In Sec.~\ref{sec:disc}, we discuss the physical origin of the reduced star formation activities and their prevalence in the overdensities. In Sec.~\ref{sec:sum}, we provide the summary. Where relevant, we adopt the AB magnitude system \citep{oke83,fukugita96}, cosmological parameters of $\Omega_m=0.3$, $\Omega_\Lambda=0.7$, $H_0=70\,\kms\, {\rm Mpc}^{-1}$, and the \citet{chabrier03} initial mass function (IMF).

\begin{figure*}
\centering
    \includegraphics[width=0.7\textwidth]{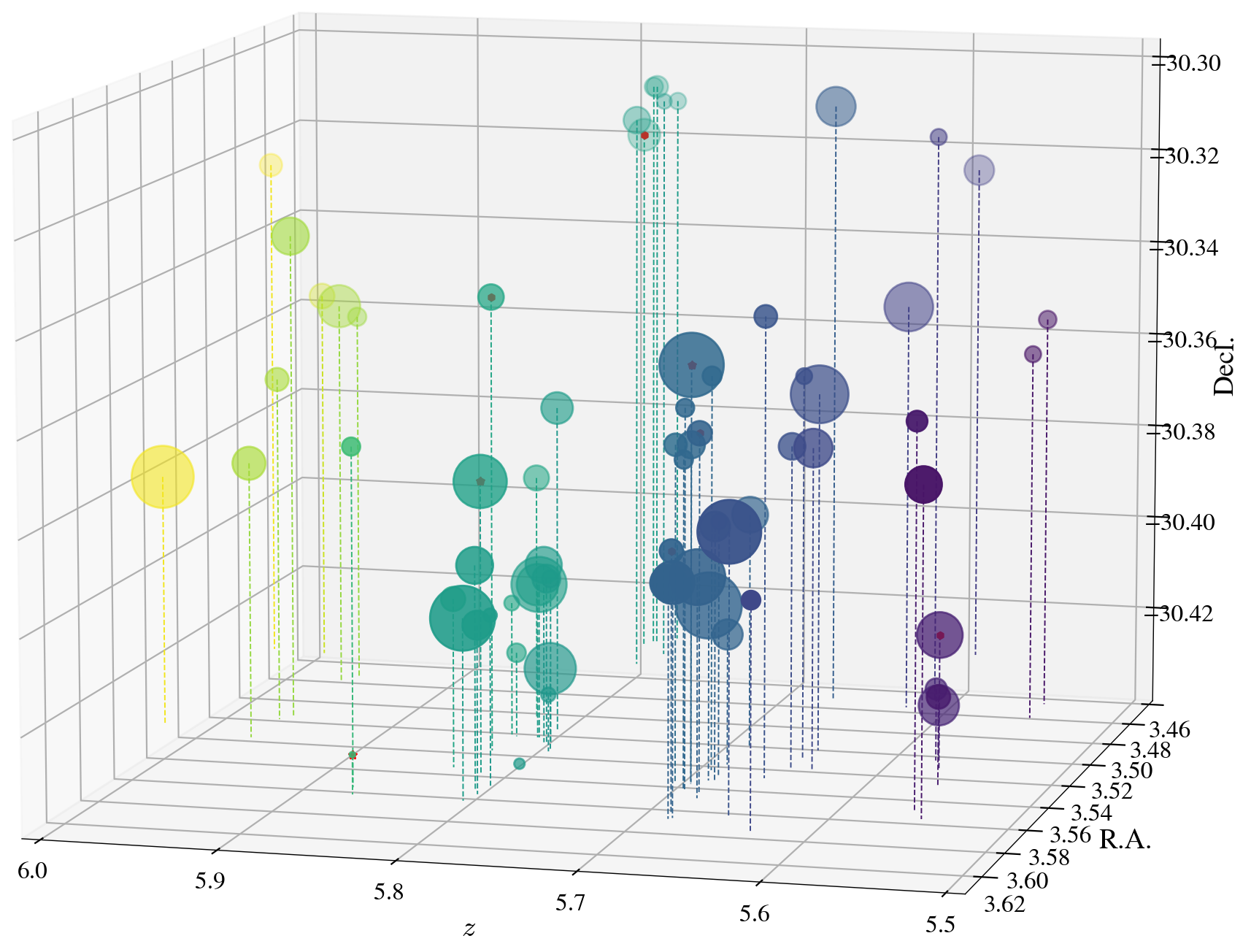}
    \includegraphics[width=0.47\textwidth]{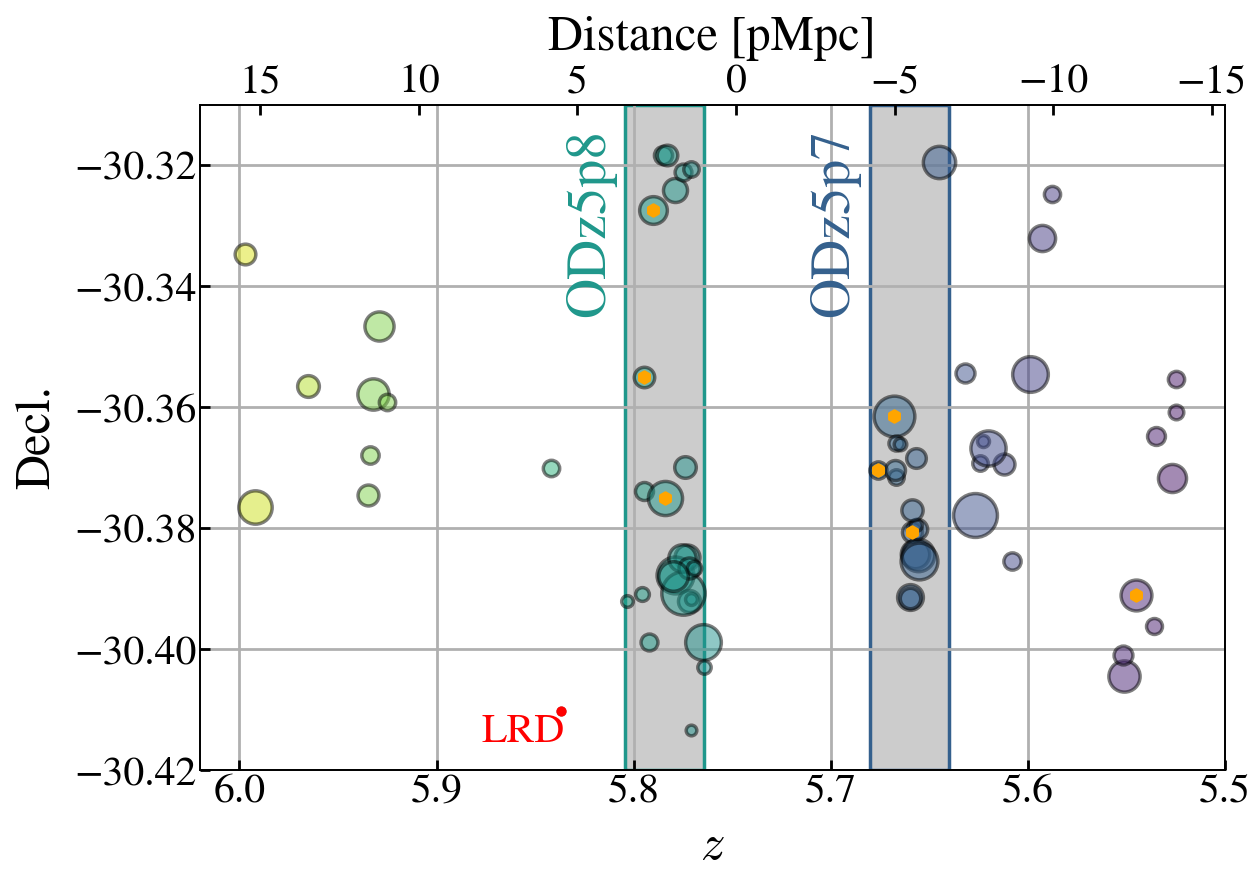}
    \includegraphics[width=0.47\textwidth]{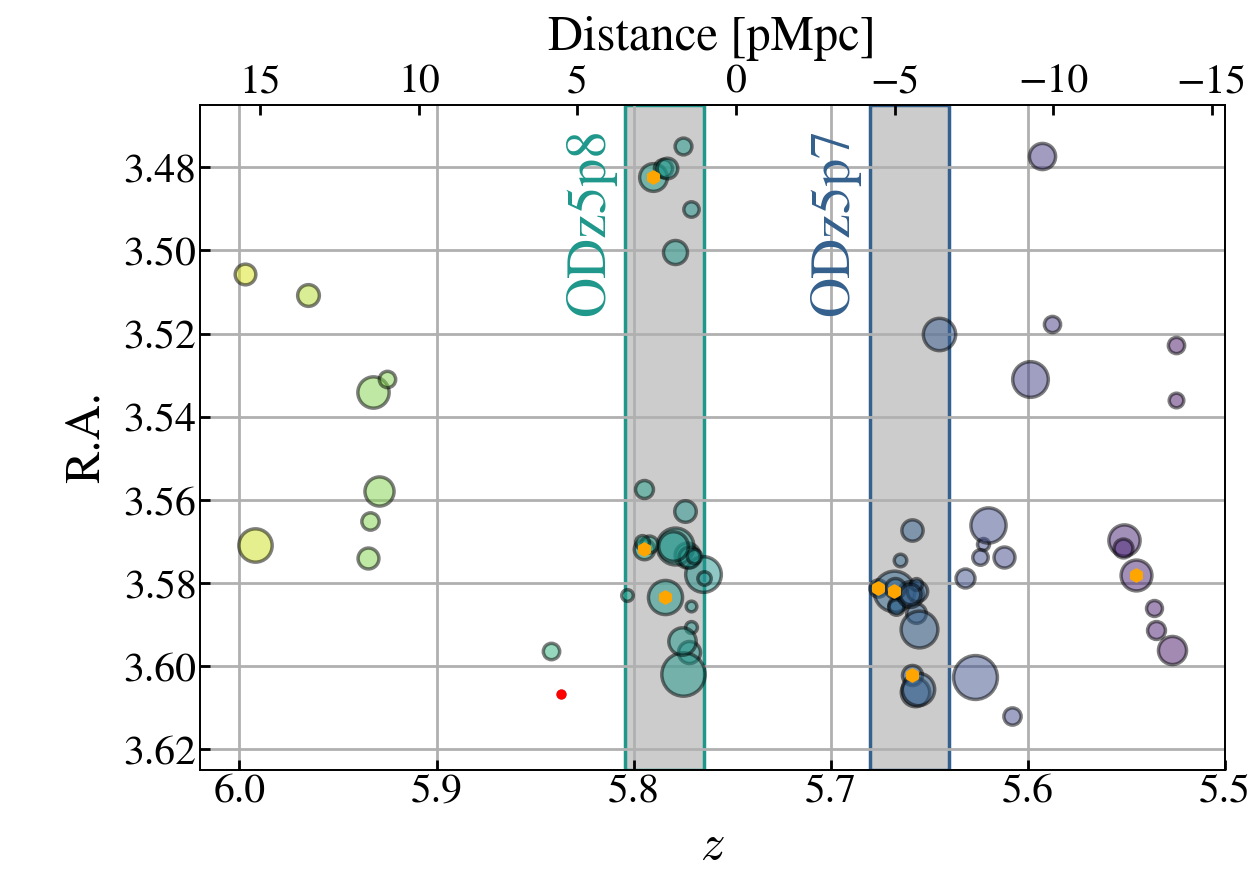}
    \includegraphics[width=0.47\textwidth]{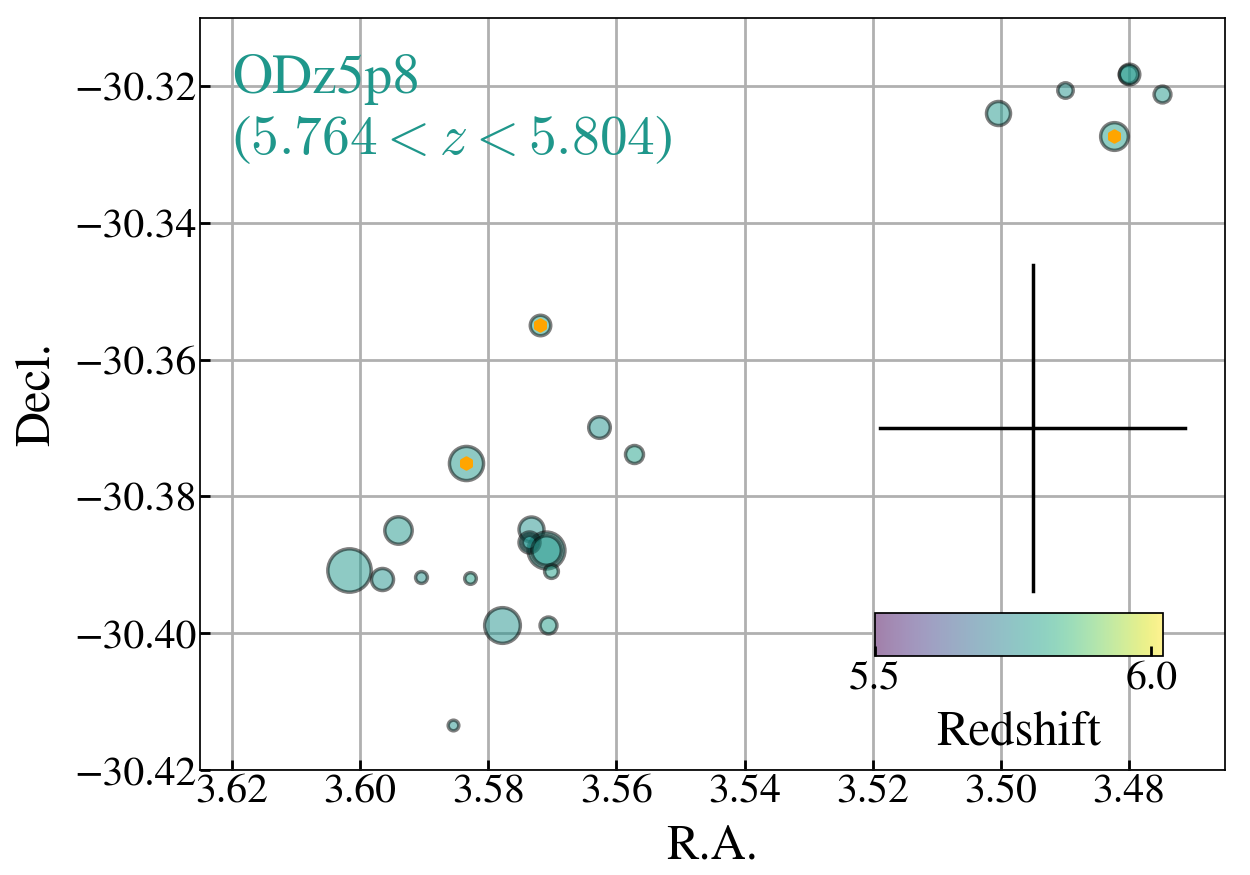}
    \includegraphics[width=0.47\textwidth]{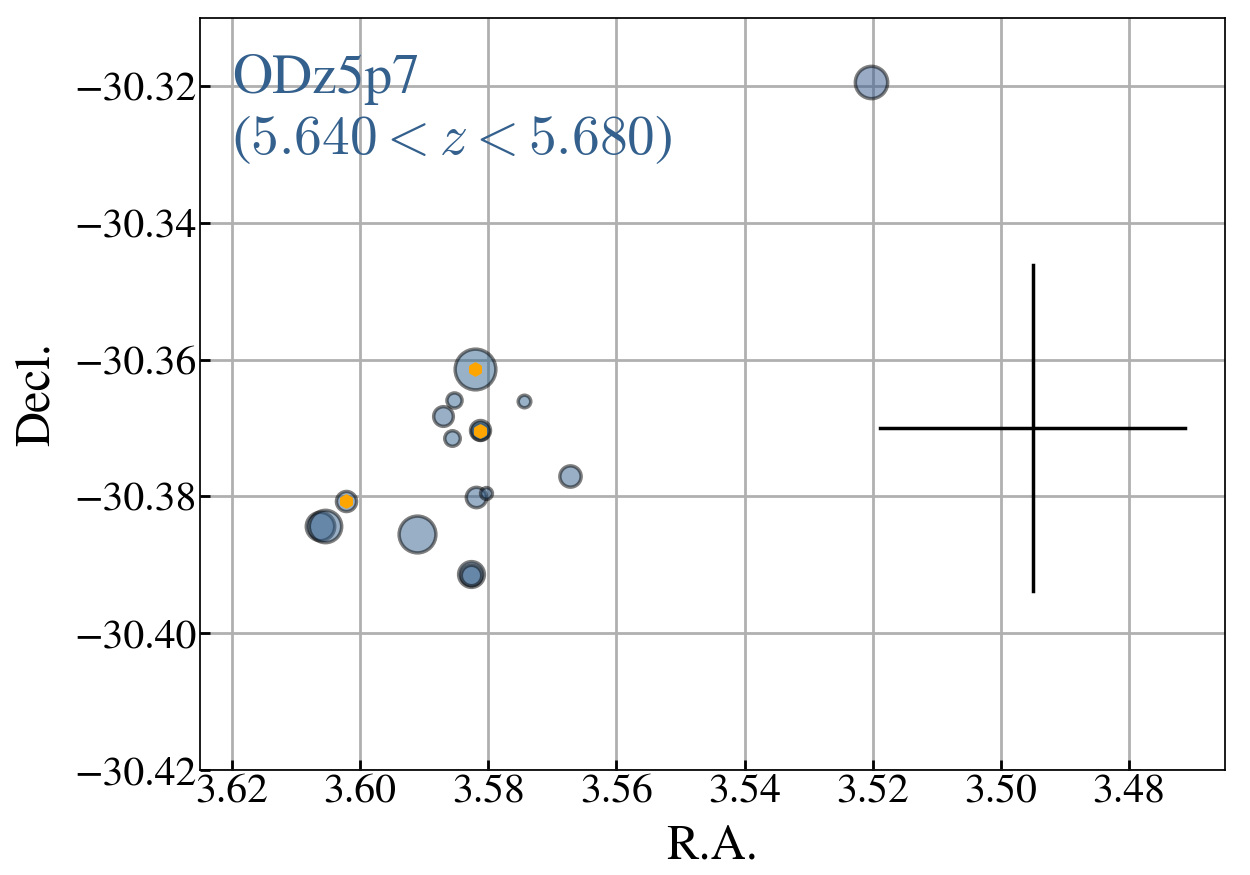}
    \caption{
 Distribution of spectroscopically confirmed galaxies at $5.5<z<6.0$ in the Abell2744 field, in a three-dimensional space (top) and two-dimensional planes (bottom four panels). Two overdense regions are identified at $z=5.68$ ({A2744-}ODz5p7) and $5.78$ ({A2744-}ODz5p8). The data points are color-coded by redshift and scaled by stellar mass. The galaxies showing weak \ha+\nii\ emissions are identified (Sec~\ref{sec:ew}) and marked separately (orange hexagons). Note that lens magnification has been corrected and the sources are shown in their source plane coordinates. Scale bars for $\sim1$\,pMpc are shown in the bottom two panels.
 The isolated red point at $z=5.84$ is a ``Little Red Dot" (Sec.~\ref{sec:member}).
    }
\label{fig:mosaic}
\end{figure*}

\section{Data}\label{sec:data}
We briefly describe the JWST/NIRCam+NIRISS imaging and NIRSpec spectroscopic data used in this work. Our target field in this paper is in the sightline of Abell~2744 (hereafter A2744), a massive cluster of galaxies at $z=0.308$ (00:14:18.9, -30:23:22). 

\subsection{JWST/NIRCam and NIRISS Imaging}\label{sec:nircam}

The field has a wealth of NIRCam data sets taken in multiple programs in Cycles 1 and 2, including GLASS-JWST \citep[ERS1324;][]{treu22}, UNCOVER \citep[GO2561;][]{bezanson22}, DD2756 \citep{chen22}, GO2883 (PI F. Sun), All the Little Things \citep[ALT; GO3516;][]{naidu24}, GO3538 (PI E. Iani), BEACON \citep[GO3990;][]{morishita24beacon}, and Medium Bands, Mega Science \citep[GO4111;][]{suess24}. {The field also has partial coverage by NIRISS imaging taken as part of the programs ERS1324 and GO2561.} We follow the same steps for JWST imaging reduction and photometry as presented by \citet{morishita23}. Briefly, we retrieved the level1 ({\tt uncal.fits}) NIRCam and NIRISS imaging data from MAST and processed with the official JWST pipeline\footnote{https://github.com/spacetelescope/jwst}. The reduction process includes extra steps to minimize artifacts, such as extra snow ball masking and $1/f$-noise subtraction. The images are then combined in each filter and aligned to a reference image, here we use the publicly available image produced by the GLASS-JWST team \citep{merlin22,paris23}. The final images are then resampled in the same pixel grid of the pixel size of $0.\!''0315$. We utilize the HST ACS and WFC3-IR mosaic images publicly available by the GLASS-JWST team, which combined data from multiple HST programs \citep{postman12,lotz17,kelly18,steinhardt20}. {The filters included in our analysis and their typical limiting magnitudes ($5\,\sigma$, for point source) are: HST-F435W (28.7), F606W (28.1), F775W (28.3), F814W (27.7), F105W (27.5), F125W (27.5), F140W (28.2), F160W (27.1), NIRCam-F070W (28.5), F090W (28.9), F115W (28.7), F140M (27.9), F150W (28.8), F162M (27.9), F182M (28.1), F200W (28.9), F210M (28.0), F250M (28.0), F277W (29.1), F300M (28.1), F335M (28.1), F356W (29.1), F360M (28.2), F410M (28.6), F430M (27.7), F444W (28.9), F460M (26.8), F480M (27.2), NIRISS-F115W (28.8), F150W (28.7), F200W (28.6), F356W (28.8), F430M (27.6), F444W (28.5), F480M (27.0).}

We identified sources in the IR detection image (F277W+F356W+F444W stack) using SExtractor \citep{bertin96}, and measured fluxes on the mosaic images {(matched to the F444W PSF)}, with a fixed aperture of radius $r=0.\!''16$\,. The measured fluxes are calibrated using the method introduced in \citet{morishita23}. The magnification by the foreground cluster is corrected for each source using the latest magnification model by \citet{bergamini23,bergamini23b}.

In addition to previously confirmed spectroscopic sources (details in Sec.~\ref{sec:msa}), we identify candidate galaxies at $z\sim5.7$ to be inspected in our NIRCam { Wide Field Slitless Spectroscopy (WFSS)} analysis in Sec.~\ref{sec:wfss}. For this, we preselect the sources using the Lyman break dropout method \citep{steidel98}, followed by a stringent photometric redshift selection, as adopted in previous studies \citep[][]{morishita23c}. We apply the following selection cut:

$$S/N_{\rm F444W} > 5 $$
$$S/N_{\rm <1216} < 3$$
$$5.4<z_{\rm phot}<6.2,$$
where the blue filters (rest-frame wavelength of $<1216\,{\rm \AA}$ i.e. F070W and bluer) are used for non-detection, and $z_{\rm phot}$ is the peak redshift {(i.e. where the redshift likelihood is maximized)}, obtained with phot-z code {\tt eazy} \citep{brammer08}.

We utilize the JWST+HST photometric data to infer the spectral energy distribution (SED) of spectroscopically confirmed galaxies, by using SED fitting code {\tt gsf} \citep[ver1.8.5;][]{morishita19}. The SED-based star-formation rate is calculated with the rest-frame UV luminosity ($\sim1600\,\rm{\AA}$) using the {\tt gsf}-determined SED model. The UV luminosity is then corrected for the inferred dust attenuation. The attenuation-corrected UV luminosity is converted to star formation rate (SFR) via the relation by \citet{kennicutt98_sfr}:
\begin{equation}
    {\rm SFR\,[M_\odot\,yr^{-1}]} = 1.4 \times 10^{-28} L_{\rm UV}\,[{\rm erg\,s^{-1}\,Hz^{-1}}].
\end{equation}
The estimated SFR is then corrected for the Chabrier IMF, by multiplying by a factor of 0.63 \citep[][]{madau14}. The properties derived from the SED analysis are summarized in Table~\ref{tab:phys}. {We note that the scaling factor was originally derived for populations of solar metallicity and may differ for lower-metallicity galaxies \citep[e.g.,][]{shapley23b}. However, because precise metallicity measurements are not available for most galaxies presented here, we coherently adopt the aforementioned factor. We find that for our samples the SFRs calculated this way are consistent with those averaged over the last 10\,Myr of the best-fit star formation history derived by {\tt gsf}. {We found that SFRs directly derived from the observed \ha\ fluxes are on average $\sim0.2$\,dex lower than UV-based ones but show large scatters, with a standard deviation of 0.4\,dex, after correcting for aperture loss by the MSA shutter \citep[e.g.,][]{morishita23c}. The difference can also be attributed to various factors, such as star formation histories, the IMF, or a strong color gradient within a single galaxy (which makes the aperture correction even more complicated).}

\subsection{JWST/NIRSpec MSA}\label{sec:msa}

We utilize the spectroscopic catalog presented by \citet{roberts-borsani24}. The work processed publicly available JWST/NIRSpec Micro-Shutter Assembly (MSA) PRISM spectroscopic data in multiple fields. Spectroscopic redshift measurements were complied after identification of bright optical emission lines, such as \oiii\ and \ha, and visual inspection for any flagging sources. Readers are referred to the original work for full details of the reduction and analysis. The catalog consists of 167 sources at $z>5$ in the A2744 field, collected from three observing programs, GO2561, DDT2756, and GO3073 \citep{castellano24}. {Exposure times for individual sources are reported in Table~\ref{tab:phys}.} {The typical exposure time for our sources is 5.7\,hrs.}

For the line-flux measurements, we fit each bright line with a Gaussian model. Each line is assigned a redshift parameter. {However, the lines close in wavelength (i.e. \hb+\oiii\ doublets, \ha+\nii\ doublets) are fitted simultaneously using a common redshift parameter. The line widths are also set as a free parameter for each line, except for the \oiii-doublet lines which we set a single width parameter.} The flux continuum is estimated by fitting a second-order polynomial over the non-line wavelength range defined for each targeted line and is subtracted before the line fit. The flux contribution from the absorption of Balmer lines is corrected by using the best-fit stellar template (without emission line component) derived by the aforementioned SED modeling.

\subsection{JWST/NIRCam WFSS}\label{sec:wfss}
Unlike NIRSpec-MSA, WFSS offers unbiased identification of strong emitters without requiring pre-selection of sources. We utilize two NIRCam-WFSS spectroscopic programs that are configured with the NIRCam F356W or F335M filters (GO3516 and GO3538), which offer wavelength coverage for \hb+\oiii\ lines at the redshift range of our interest {$5.5<z<6$}, at observed $3.16$--$3.50\,\mu$m. The total exposure time ranges from 10--100\,ksec (Figure~\ref{fig:fov}; Table~\ref{tab:phys}). {The typical exposure time for our sources is 9.8\,hrs.}

We follow a similar reduction process as \citet{liu24}, which is motivated by the median filter technique introduced in \citet{kashino23}.
The reduction adopts ``line scanning'' after subtracting continuum fluxes over the wavelength range of interest for each target. This process involves a combination of the official JWST pipeline and several custom steps, as detailed below. We start by obtaining the stage-1 products (\texttt{rate.fits}) from the MAST archive. We then assign the world coordinate system (WCS) and perform flat fielding on each frame using the most recent data available from the JWST Calibration Reference Data System (CRDS). Our background subtraction process consists of two steps. First, we subtract a median background created separately for each module and pupil. To further remove any remaining background, we use \texttt{SExtractor} \citep{bertin96} for additional subtraction. We then extract the emission line by removing the continuum, which is modeled using a median filter method with a smoothing kernel. With the background and continuum-subtracted images, we are able to search for \hb+\oiii\ emission lines in photometrically selected galaxies (Sec.~\ref{sec:nircam}) by locating the 2D spectra along the dispersion direction, using the spectral tracing and grism dispersion models generated by \citet{sun23}.

Given the small wavelength range of the extracted spectra and potential contamination by other sources at a similar position in the cross-dispersion direction, we only include sources that have the pair of \hb+\oiii\ doublet lines. This means that those with a single line cannot unambiguously have their redshift determined.

\begin{figure*}
\centering
\includegraphics[width=0.88\textwidth]{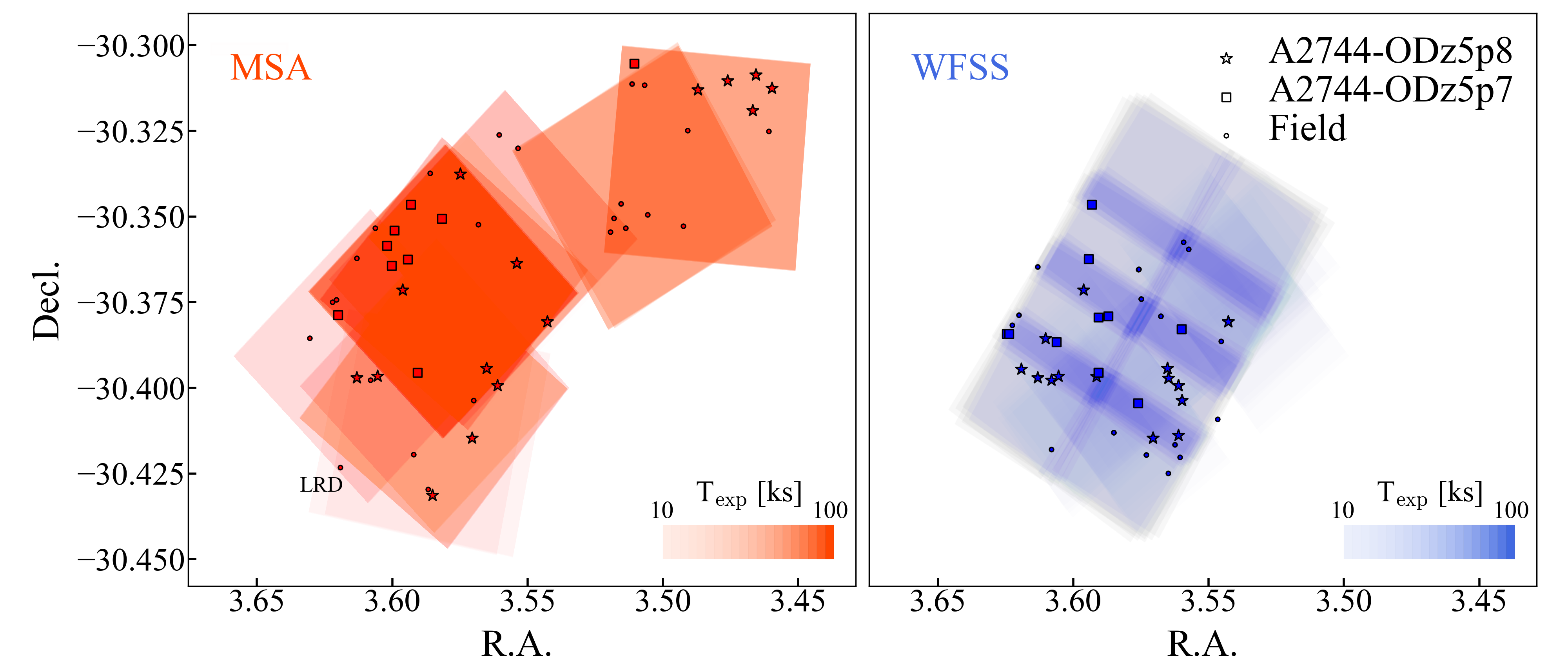}
 \caption{
Spectroscopic data coverage (MSA in the left panel and WFSS in right). Spectroscopically confirmed sources by the corresponding instrument are shown in each of the panels. The color transparency represents the total observing time {(but not necessarily does for individual MSA sources; see Table~\ref{tab:phys} for the exact exposure times)}. Note that the sources are shown in the observed sky positions, whereas Figure~\ref{fig:mosaic} shows their source-plane positions after correcting for lens magnification. 
}
\label{fig:fov}
\end{figure*}

\begin{figure}
\centering
    \includegraphics[width=0.46\textwidth]{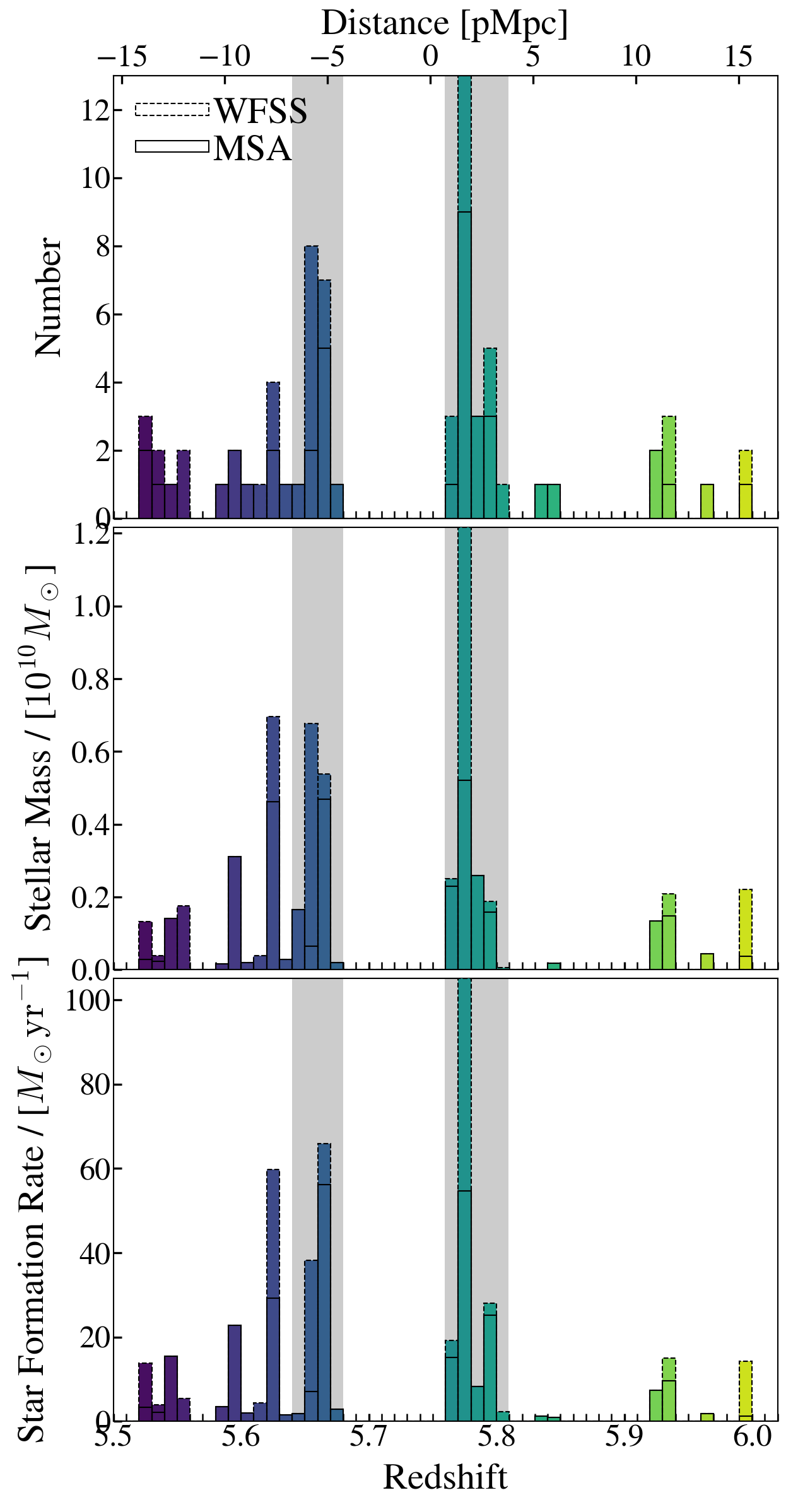}
	\caption{
 Redshift distributions of the spectroscopically confirmed galaxies in the sightline of A2744 (top). The weighted distributions by stellar mass (middle) and by star formation rate (bottom) are also shown. Solid lines are for the MSA sample and dashed lines for the WFSS sample. The redshift ranges of the two overdensites are hatched in gray. 
    }
\label{fig:zhist}
\end{figure}

\begin{figure*}
\centering
    \includegraphics[width=0.98\textwidth]{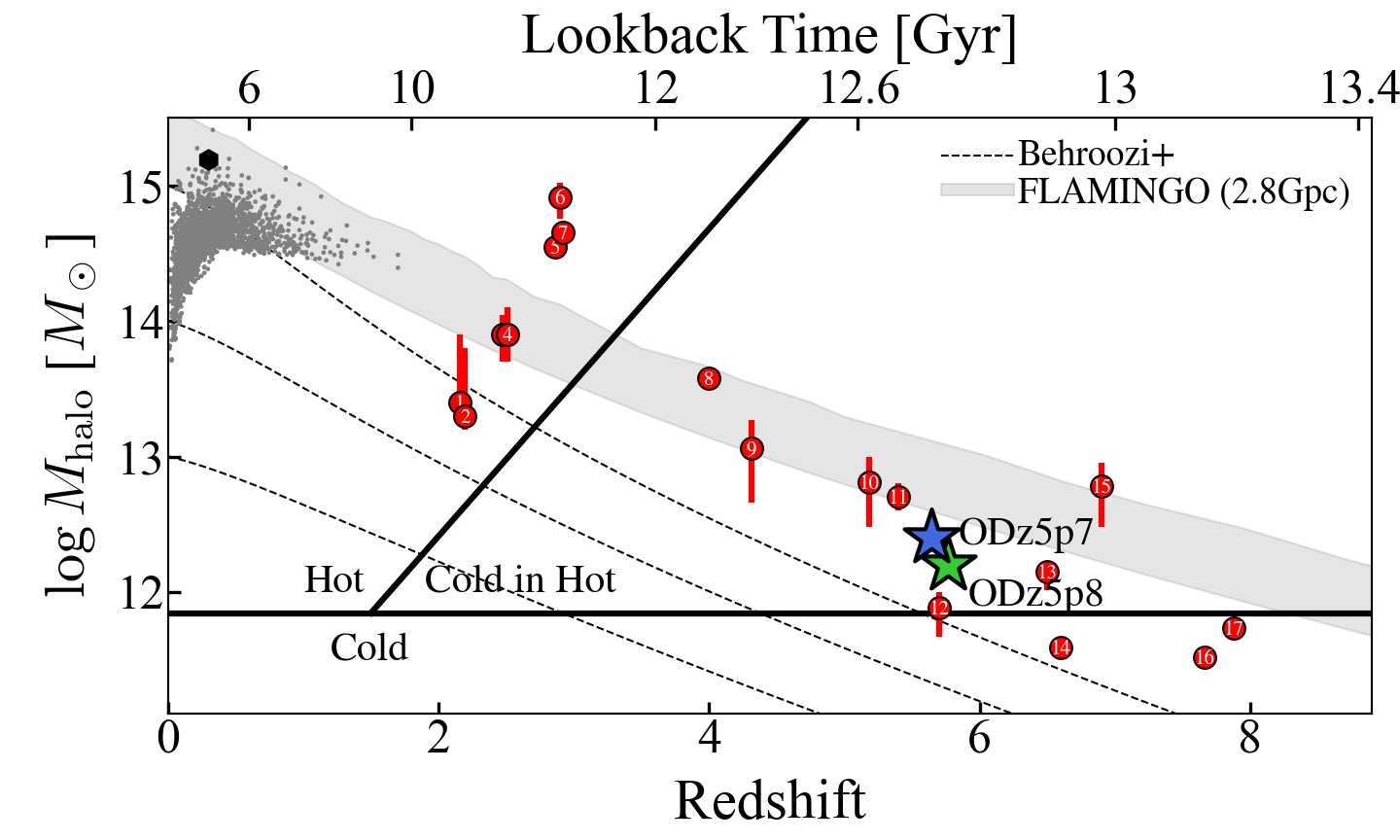}
	\caption{
Halo masses of the two discovered overdensities ({A2744-}ODz5p7 and {A2744-}ODz5p8; star symbols), along with other (proto-) clusters and overdensities in the literature. 
Trajectories of massive halos ($M_{\rm h}(z=0)=10^{13}, 10^{14}$, and $10^{15}\,M_\odot$) calculated using an empirical formulae \citep[][dashed lines]{behroozi13} and of $>10^{15}\,M_\odot$ identified in the FLAMINGO simulation \citep[][hacthed regions]{lim24} are shown. Data points at $z>2$ are: 
1 and 2 \citep{polletta21,koyama21}, 3 \citep{casey15}, 4 \citep{wang16}, 5 \citep{venemans07}, 6 \citep{cucciati14}, 7 \citep{venemans07}, 8 \citep{oteo18}, 9 \citep{miller18}, 10 \citep{sun23}, 11 \citep{helton24}, 12 \citep{harikane19}, 13 \citep{chanchaiworawit19}, 14 \citep{harikane19}, 15 \citep{arribas23}, 16 \citep{laporte22}, 17 \citep{morishita23b}.
Data points at $z<2$ (gray) are SPT clusters presented in \citet{bocquet18} and X-ray-SZ clusters in \citet{tarraio19}. The three different gas cooling regimes (cold, hot, and cold in hot) predicted by \citet{dekel06} are shown. In the hot regime, cold flows are expected not to reach the central galaxy due to shock heating. The halo mass of A2744OD7p9 has been updated from the one estimated in \citet{morishita23b} to $5.8\pm0.3\times10^{11}\,M_\odot$ \citep{morishita25}, by including three newly confirmed sources \citep{hashimoto23,chen24}. The total mass of the cluster Abell~2744 estimated in \citet{finner24} is also shown (black hexagon). 
}
\label{fig:Mhalo}
\end{figure*}

\section{Discovery of Two Overdensities of Galaxies at $z\sim5.7$}\label{sec:member}

Figure~\ref{fig:mosaic} shows the distribution of spectroscopically confirmed sources at $5.5<z<6.0$ in the A2744 field. Remarkably, we find two concentrations of galaxies at $z\sim5.66$ and $z\sim5.78$. By taking a redshift window of $\Delta_z = 0.04$, which corresponds to $\sim 2$\,pMpc (similar to the probed projected area), the two overdensities consist of \ntotza\ and \ntotzb\ member galaxies, respectively. Despite being in the same sightline, the physical separation of the two overdensities is $\sim6$\,pMpc along the redshift space. As such, in what follows we proceed our analyses by treating the two overdensities as isolated from each other, rather than as a single large-scale structure. We hereafter refer to these overdensities as {A2744-}ODz5p7 ($5.640<z<5.680$) and {A2744-}ODz5p8 ($5.764<z<5.804$). For simplicity, we refer to the remaining sources as ``field" samples ($N=$\,\ntotfld).

Most of the member galaxies are concentrated within the area of projected $R\sim0.5$\,pMpc. There are a few galaxies at a further distance, $\sim3$\,pMpc from the primary concentration. We note, however, that the empty region between the primary and secondary concentrations is, if not completely, likely due to the lack of WFSS coverage (Figure~\ref{fig:fov}). As such, mass and star formation rate measurements of the overdensities themselves in what follows likely serve as lower limits.

{The observed sizes of the two overdensities correspond to the typical size of the protoclusters of similar redshifts seen in numerical simulations \citep[e.g.,][]{chiang17}. We note that the redshift window was centered so that the number of member galaxies is maximized in each overdensity.} 

We estimate the overdensity factor, $\delta=(n-\bar{n})/\bar{n}$, which represents the excess of surface number density from the field average, by following the same procedure as in \citet{morishita23b}. In each overdensity, we use $R=0.5$\,pMpc and the redshift range as defined above. For field reference, we used the luminosity function at $z\sim6$ presented in \citet{bouwens21} and integrated it down to $M_{\rm UV}\sim-18$\,mag, approximately the lower $1\,\sigma$ from the mean observed $M_{\rm UV}$ in each concentration. For the area within $R=0.5$\,pMpc, {we expect $N\sim35$ galaxies in average fields. With the reference number, we find $\delta = 4.1_{-1.4}^{+5.2}$ and $6.6_{-2.4}^{+9.1}$ for {A2744-}ODz5p7 and {A2744-}ODz5p8, respectively, where the uncertainties quoted represent Poisson noise. Similarly, we calculate the overdensity factor of the field sample and find $\delta=-0.2_{-0.1}^{+0.1}$, which is slightly underdense but consistent with the expectation of the luminosity function.}

Figure~\ref{fig:zhist} shows the redshift distributions of galaxy counts, the total stellar mass, and the total star formation rate in each redshift bin. Both overdensities have similar total stellar masses ($\sim2\times10^{10}\,M_\odot$) and total star formation rates ($\sim200\,M_\odot {\rm yr^{-1}}$). 

The halo mass at the time of the observed redshift is estimated by integrating the halo mass of individual member galaxies, where the individual halo mass is calculated by the empirical $M_h$--$M_{\rm UV}$ relation at $z\sim6$ in \citet{mason22}. We estimate it to be $\log M_h / M_\odot \approx 12.0$ and $\approx12.2$ for {A2744-}ODz5p7 and {A2744-}ODz5p8, respectively. Instead, when the empirical $M_h$--$M_*$ relation at $z\sim6$ in \citet{behroozi19} is adopted, we obtain $\log M_h / M_\odot \approx 12.4$ and $\approx12.6$. The increased halo mass by the latter method is likely due to the fact that our galaxies are not very young, which is instead assumed for the $M_h$--$M_{\rm UV}$ relation, leading to a reduced $M_h$ value for a given $M_{\rm UV}$.

Both the estimated $\delta$ and the halo masses suggest that each overdensity is expected to evolve into a system of $\simgt10^{14}\,M_\odot$ by $z\approx0$ \citep[e.g., Figure 5 in][]{toshikawa14}. Figure~\ref{fig:Mhalo} shows the halo masses of the two overdensities along with the measurements in the literature. Predicting the mass of a descendant system generally involves uncertainties, especially when the field of view (FoV) is limited \citep[see Sec.~4.2 of][]{morishita23b}. Nevertheless, comparisons with the mass growth trajectories of simulations \citep{lim24} suggest that these two overdensities would have evolved into massive systems by $z\sim0$.

We find no excess of \lya\ emitters in either of the overdensities. We find one \lya\ emitter within {A2744-}ODz5p8 (of 16 having the wavelength coverage) and none within {A2744-}ODz5p7 (of 9), whereas two are found in the field sample (of 24). 

Lastly, one of the spectroscopically confirmed sources, ID43506 at $z=5.84$, exhibits a v-shape spectrum, a characteristic feature of the recently discovered ``Little Red Dot" population \citep[e.g.,][]{furtak23,matthee23,greene24}. Furthermore, the spectrum shows a larger line width in \ha\ and \hb\ {($\sigma\approx800$\,km/s), from what the prism resolution can afford i.e. $\sim300$--600\,km/s for the observed \hb\ and \ha\ wavelengths).}
This source appears to be isolated from other galaxies in the field (Figure~\ref{fig:mosaic}), {while it is located at the edge of the spectroscopic coverage, requiring further comparable coverage for the definitive conclusion}. We found that the estimated stellar mass and star formation rate for this source are uncertain, as the observed fluxes could be dominated by AGN. As such, we exclude this source in the following analysis.

\section{Spectroscopic Properties of Member Galaxies}\label{sec:phys}

The sensitivity and wavelength coverage of NIRSpec MSA spectra allow us not only to confirm spectroscopic members, but also to investigate the physical properties of individual galaxies. In this section, we focus on the MSA samples for a further characterization of their properties. For simplicity, hereafter we refer to galaxies in one of the two overdensities as overdensity samples, and others as field samples. 

\begin{figure}
\centering
    \includegraphics[width=0.49\textwidth]{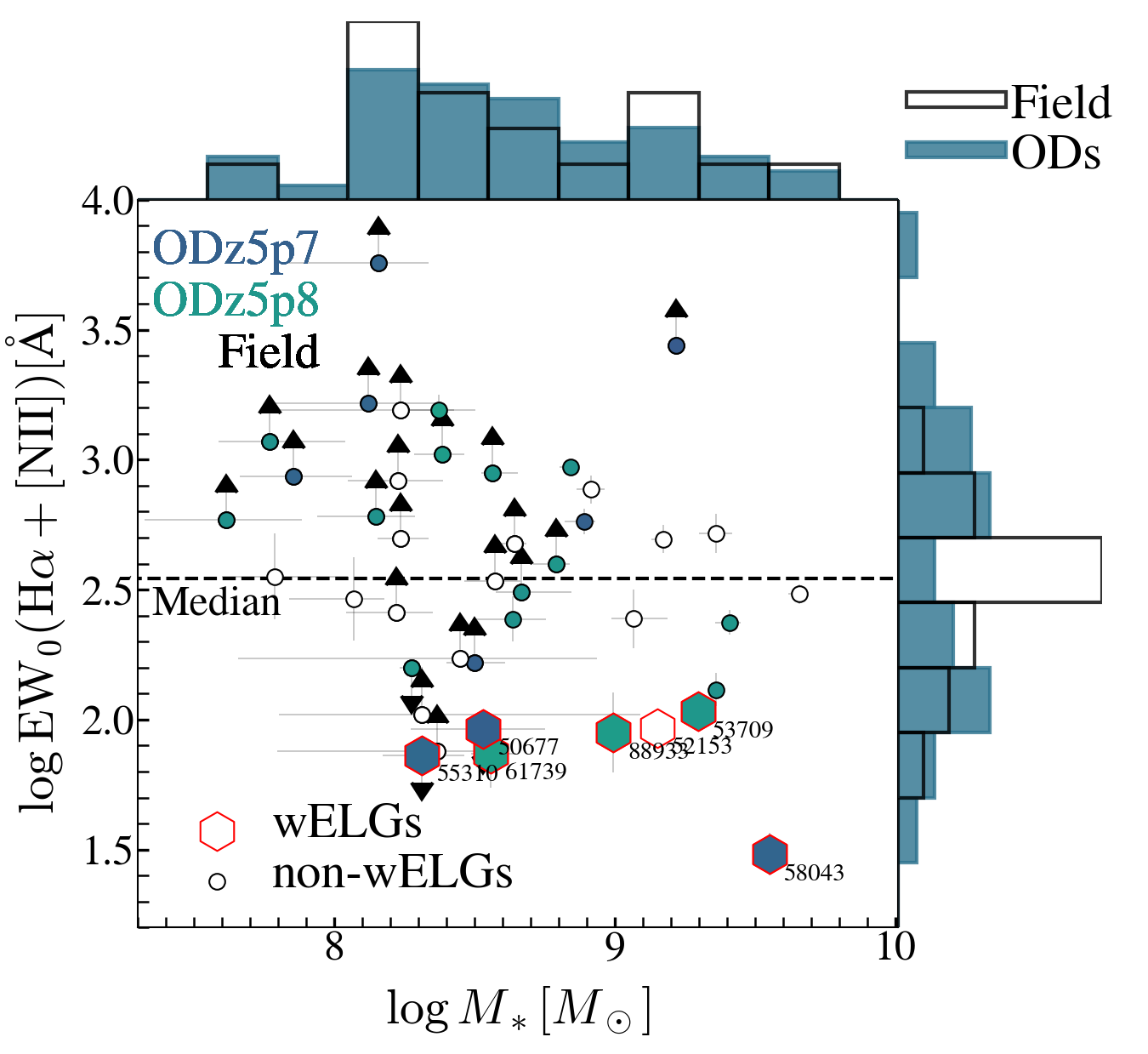}
	\caption{
 Distribution of \ha+\nii\ rest-frame equivalent width measurements of the overdensity samples (filled symbols) and the field samples (open symbols). 2-$\sigma$ lower limits are shown for those with the continuum undetected (arrows). Those with \ewh\,$\simlt100\,{\rm \AA}$ are defined as weak emission line galaxies (wELGs; hexagons). IDs are shown only for wELGs. A high fraction (6/7) of wELGs are found in {the two overdensities identified here}. The median value measured for the entire sample is shown (horizontal dashed line). 
 }
\label{fig:ew}
\end{figure}

\begin{figure*}
\centering
    \includegraphics[width=0.99\textwidth]{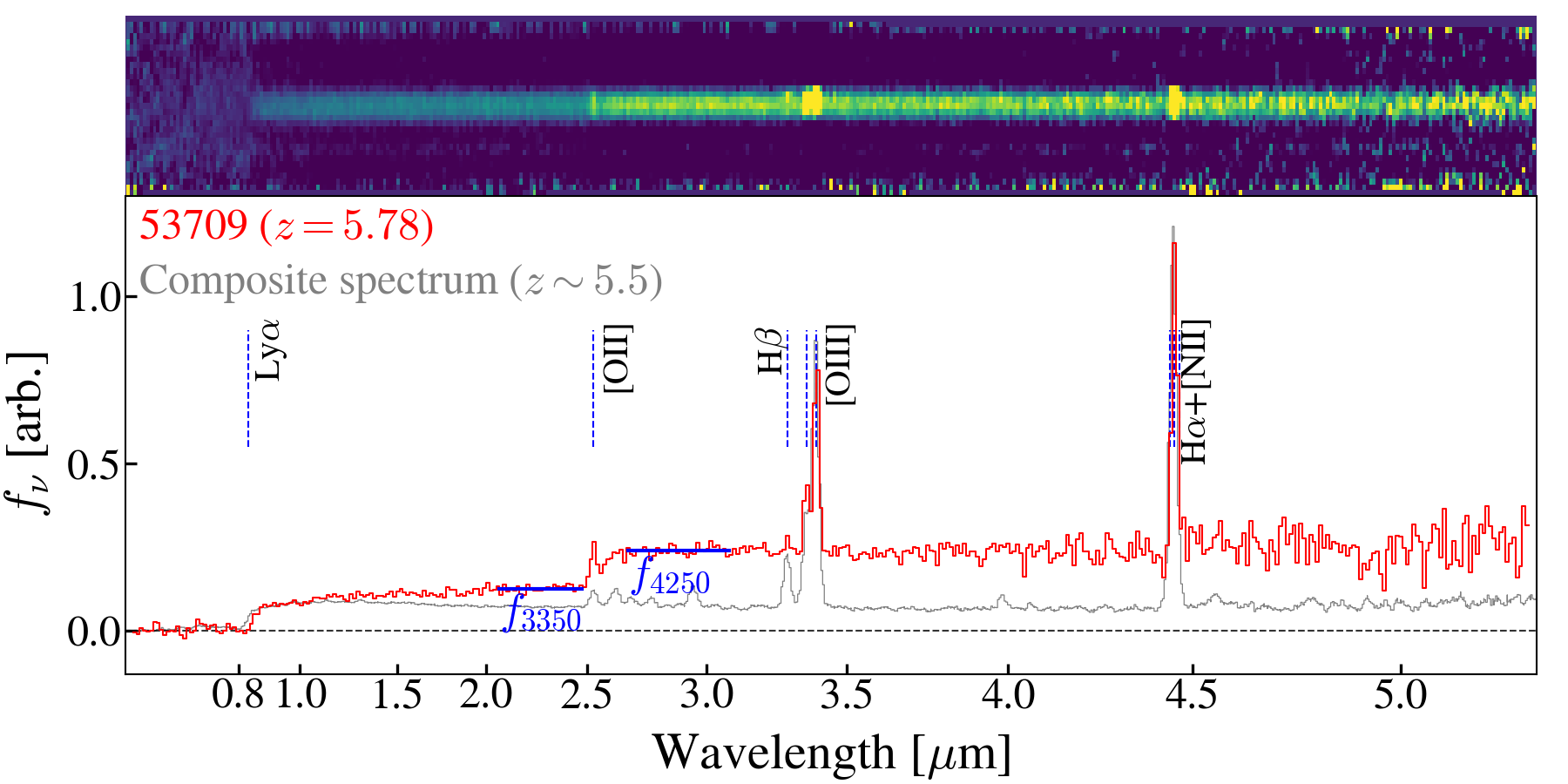}
	\caption{
 One of weak emission line galaxies (wELGs; Sec.~\ref{sec:ew}). This example showcases a prominent continuum ($4000$\,\AA+Balmer) break, measured by the ratio $f_{4250}/f_{3350}$, indicating the presence of evolved ($\approx100$\,Myr) stellar populations. 
 Compared is the average spectrum of $z\sim5.5$ galaxies (gray line, {redshifted to $z=5.78$}) from \citet{roberts-borsani24}.
}
\label{fig:spec_wel}
\end{figure*}
\subsection{\ha+\nii\ Emission Line Equivalent Width}\label{sec:ew}

For the redshift range of the two overdensities, \ha+\nii\ emissions lines fall within the spectral range of NIRSpec PRISM. The spectral resolution at the relevant wavelength ($\sim4.4\,\mu$m) is $R \sim 200$, which barely resolves the \ha\ from the \nii\ doublets. As such, while we fit the line complex with a three-component Gaussian model (Sec.~\ref{sec:msa}), we only use the integrated flux of the three lines in the following discussion. Among the NIRSpec MSA sample, we could obtain a reliable EW measurement in {43} galaxies. We note that the WFSS samples are not covered with their \ha\ or \nii\ lines, and thus they are excluded from the discussion below.

The measured rest-frame equivalent widths, \ewh, are shown in Figure~\ref{fig:ew}. For those with the underlying continuum undetected or uncertain, we quote a lower limit by using the RMS of the spectrum in the relevant wavelength range. For those with the continuum detected but emission lines undetected, we quote an upper limit. 

Most of the sample galaxies range between $\sim200$ and $2000\,{\rm \AA}$, with a median value of $\sim300\,{\rm \AA}$, which is consistent with the measurements at similar redshifts by \citet{roberts-borsani24}. Among those, we find nine galaxies showing a significantly reduced value, $\simlt100\,{\rm \AA}$, which we hereafter refer to as weak emission line galaxies, or wELGs. The identified wELGs have a relatively high mass ($\logm \simgt8.4$) compared to the entire distribution. 

Remarkably, all except one wELGs are found within one of the two discovered overdensities. Among the MSA samples, the fraction of wELGs in the overdensities is {3/9 ({A2744-}ODz5p7) and 3/16 ({A2744-}ODz5p8), whereas it is much smaller in the field sample (1/18).} It is noted that the dominance of wELGs in the overdensities is unlikely due to the selection effect (see also Sec.~\ref{sec:sele}). While the WFSS samples are not included in the statistics, adding those as non-wELGs would give fractions of 3/17, 3/25, and 1/29, for {A2744-}ODz5p7, {A2744-}ODz5p8, and the field sample, respectively.

In addition, we investigate the fraction of wELGs in other fields of \citet{roberts-borsani24}. Among 141 sources at $5.5<z<6.0$ that are successfully fitted, we find eight sources that satisfy \ewh\,$<100\,{\rm \AA}$, which gives a similar ratio of wELGs to our field sample. This confirms that the field sample defined in this study does not differ significantly from representative samples at the same redshift. 

From the comparison of the fractions of wELGs in the two field samples, it is suggested that wELGs are more ubiquitous in overdensities. In fact, the only wELG in the field sample (ID52153) is located in a less pronounced overdensity, a concentration of four galaxies, which further supports the hypothesis.

We find a larger scatter observed in the \ewh\ measurements of the over-density samples than in the field samples (Figure~\ref{fig:ew}). Not only wELGs, the overdensity sample consists of a few emitters at the {\it high} EW end ($\simgt1000\,{\rm \AA}$) too---{at least 6 of the overdensity samples ($>14\,\%$), whereas 1 in the field sample ($>3.4\,\%$).} This suggests that the star formation activity of galaxies in overdensities is likely less constrained than in the field sample, implying the presence of physical processes that are more common in the former environment. We revisit this in Sec.~\ref{sec:disc}.

\subsection{Strength of the 4000\,\AA+Balmer Break}\label{sec:bb}

The spectrum of one of the identified wELGs is shown in Figure~\ref{fig:spec_wel}. Besides the modest star formation activity measured in their emission lines, we find that wELGs generally show a pronounced continuum break at rest-frame $4000$\,\AA, a characteristic features of evolved galaxies after having a short starburst \citep[][]{dressler83,dressler99}. Such a characteristic break is not seen in the majority of sources at similar redshifts. In Figure~\ref{fig:spec_wel}, we show the composite spectrum of $z\sim5.5$ galaxies presented in \citet{roberts-borsani24} for comparison. Beside the pronounced continuum break, this example wELG shows a redder UV slope. From these differences in the two spectra, it is suggested that wELGs are distinct from the representative galaxies at similar redshift.

To quantify the difference, we measure the strength of the continuum break by taking the ratio of mean flux density in both red and blue sides, \De\,$ = f_{\nu, 4250}/f_{\nu, 3350}$\footnote{The subscript, $e$, represents ``extended," to distinguish from the conventional $D4000$ (or $D_n 4000$) index, which is defined within narrower wavelength ranges.}. The average flux densities are measured in $f_\nu$, over the rest-frame wavelength range of $3050< \lambda / {\rm \AA} <3650$ and $3950 < \lambda / {\rm \AA} < 4550$, respectively, effectively capturing both Balmer and $4000$\,\AA\ breaks. The wavelengths of any emission lines, such as \hg\ and \hd, are masked out from the continuum flux estimate.

The \De\ index can be securely measured upon the significant detection of continuum fluxes in both wavelength windows. Among 49 MSA sample, we could successfully measure the strength for 30 galaxies within the uncertainty of $<0.1$\,dex. Figure~\ref{fig:bb} shows the distribution of those break measurements as a function of light-weighted age derived from our SED analysis. 

We find that wELGs have a systematically stronger Balmer break than the non-wELG samples. This is partially due to the lack of sources of $\log t / {\rm Gyr} \simlt1.6$ in the wELG samples, the regime which is dominated by young, strong emission line galaxies. More importantly, for a given age wELGs exhibit a relatively stronger break than non-wELGs. Also shown in Figure~\ref{fig:bb} is a theoretical track from stellar population synthesis, generated with {\tt fsps} \citep{conroy09fsps}. We find that all wELGs follow the track of a single, short-burst model (i.e. SSP), whereas the majority of non-wELGs follow the track of a constant star formation model. This indicates that the origin of strong Balmer break in wELGs is likely attributed to their shorter duration of star formation, instead of wELGs being formed earlier than non-wELGs.

It is noted that dust attenuation could mildly increase the \De\ measurement too. However, most of our sample galaxies have an extinction {of $A_V<0.5$\,mag} (Table~\ref{tab:phys}). We find no statistical difference in dust attenuation between the field sample and the overdensity members, suggesting that dust attenuation by itself does not contribute to the observed offset seen between the two samples.

There are a few non-wELGs that exhibit relatively strong $4000$\,\AA\ break, $\simgt1.5$. One of them, ID53527, shows evidence of line broadening in its emission lines {($\sigma\approx1000$\,km/s)}, suggesting the presence of an AGN or a galactic-scale outflow. The other two show no clear evidence of AGN. We discuss potential contributions of AGN to the shutdown of star formation in Sec.~\ref{sec:agn}. We note that these non-wELGs are all found within the discovered overdensities.

\begin{figure}
\centering
    \includegraphics[width=0.48\textwidth]{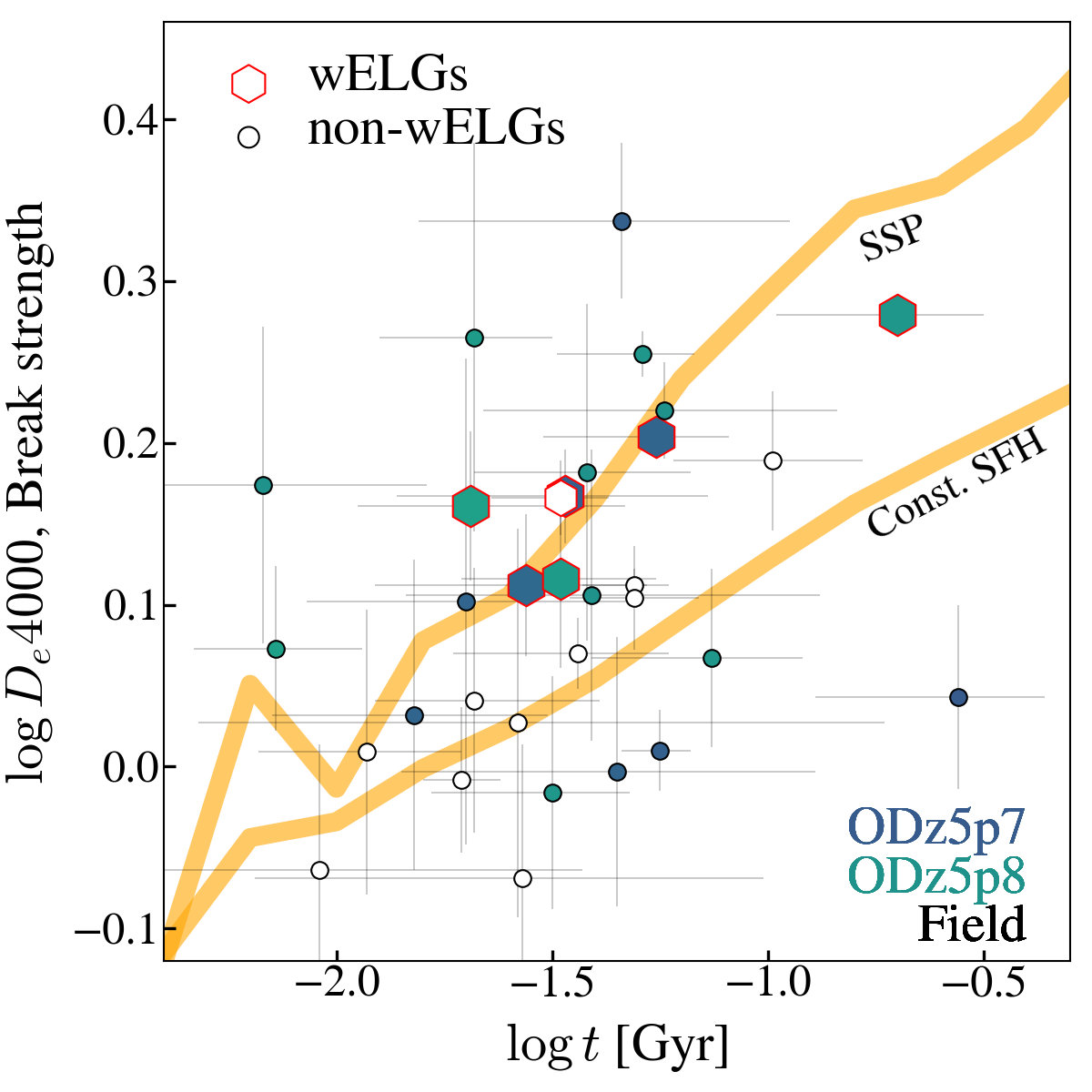}
	\caption{
 NIRSpec samples. Only those with the $D_e$4000 break securely measured (with the uncertainty of $<0.1$\,dex) are shown. Galaxies with a reduced \ha+\nii\ (i.e. wELGs; hexagons) exhibit a stronger continuum break strength than non-wELGs for a given light-weighted age ($t$, derived in the SED analysis in Sec.~\ref{sec:nircam}). Two evolutionary tracks, single burst (SSP) and constant star formation models of 10\,\%\ solar metallicity ($0.1\,Z_\odot$), are shown (orange solid lines). The wELGs follow the SSP track, indicating that they experienced a short-duration star formation. {The stellar population tracks would shift for $\sim0.08$\,dex toward the positive $\log D_e4000$ direction when either the solar metallicity or the dust attenuation of $A_V=0.5$\,mag is assumed.} 
 }
\label{fig:bb}
\end{figure}

\section{Discussion}\label{sec:disc}

\subsection{On the selection completeness of our sample}\label{sec:sele}
Our study is limited by incompleteness and pre-selection of sources for the MSA observations. While NIRCam WFSS offers an unbiased identification of emitters, it does not cover the wavelengths of \ha+\nii\ or $4000$\,\AA\ break, which are the key measurements in our discussion in the following sections. The WFSS samples, on the other hand, are found primarily within the two overdensities. In fact, as seen in Sec.~\ref{sec:ew}, the fraction of wELGs in the two overdensities is still significantly higher than in the field sample, regardless of assumptions that all WFSS are wELGs or non-wELGs. As such, our conclusion remains the same regardless of the limitation.

We also note that our selection relies on the detection of strong emission lines, and is likely incomplete for quenched galaxies. The MSA sensitivity limit for \ha+\nii\ equivalent width measurements is $\sim 10$--$50$\,\AA. However, among our photometric samples that were selected in Sec.~\ref{sec:nircam} ($N=67$), we find no galaxies that are classified as passive e.g. on the $UVJ$ color diagram, nor $\logm \simgt10$ galaxies. As such, in the following discussion we assume that the identified sources well represent the majority of the member galaxies in the two discovered overdensities.

\subsection{On the High Observed Frequency of Evolved Galaxies in Overdensities}\label{sec:}

In previous sections, we have seen that a large fraction of galaxies in the two discovered overdensities exhibit a small \ewh. While the \ha+\nii\ lines cannot completely be deblended with the NIRSpec PRISM resolution, the line ratio is typically found to be \nii/\ha\,$\sim0.1$ in star-forming galaxy populations at similar redshifts \citep[e.g.,][]{sanders23}. In the case of AGN, the contribution from \nii\ becomes larger. As such, under either of the assumptions, the measured equivalent width of \ha+\nii\ serves as a conservative indicator of ongoing star formation. 

Figure~\ref{fig:ew} shows that galaxies in the two overdensities occupy the bottom-right area, whereas we see both overdensity and field galaxies in the middle-to-high \ha+\nii\ range. The absence of sources in the bottom-left area is likely due to the incompleteness originating in the spectroscopic sensitivity. These observations indicate that star formation in {\it some} galaxies in the two overdensities is systematically suppressed compared to the field samples. 

{We aim to investigate the statistical significance of the finding. We perform the Kolmogorov–Smirnov (KS) test on the \ewh\ distributions of overdensity and field samples. For each of those with an upper or lower limit, we assign a randomly generated value permitted by their limit values. From 300 iterations of these processes, we find that the $p$-value of the KS test is $0.20_{-0.07}^{+0.03}$. Similarly, we conducted the ($k$-sample) Anderson-Darling test, finding the $p$-value of $0.09_{-0.02}^{+0.03}$. We note that the stellar-mass distributions of the two samples are similar; the $p$-value from the KS test on the mass distributions is 0.8 (also in the top panel of Fig.~\ref{fig:ew}).

These results suggest that the parent distributions of the two samples might not be identical; however, the measured $p$-values are too large to robustly reject the null hypothesis. We attribute this partially to the scattered (or bi-peak) distribution of the overdensity sample. When we limit the sample to those below the median \ewh\ value, we find a significantly lower $p$-value in both tests ($6.5\times10^{-3}$ and $3.3\times10^{-3}$).
}

We have seen in Sec.~\ref{sec:bb} that all selected wELGs are consistent with a SSP-like star formation history, whereas many non-wELG samples follow the trajectory of a continuous star formation history. On the other hand, the age distribution of wELGs and non-wELGs is found to be similar, except for the dominance of (both overdensity and field) non-wELGs at a very young age. From this comparison, it is suggested that wELGs are populations of galaxies that experienced a shorter-duration star formation but not necessarily formed earlier than the field samples. {Following the above statistical test, we find no significant evidence that the parent distributions are different for both age ($p=0.52$) and \De\ ($p=0.45$). Also, while emission line equivalent widths are often related to, e.g., specific star formation rates, we see only a weak correlation between the two parameters (with the Pearson correlation coefficient $r=0.55$).}

These findings are overall consistent with studies of galaxies in lower redshift clusters and groups, where a large fraction of post-starburst galaxies are seen \citep[e.g.,][]{dressler83,dressler99,Balogh99,paccagnella17,paccagnella19}. The physical origin of starbursts and the rapid shutdown of star formation in cluster environments have been intensively discussed, attributing to cluster-specific mechanisms such as ram pressure stripping \citep[e.g.,][]{poggianti09,vulcani20,werle22}. Their origin in early overdensities remains an open question.
Being at high redshift, our study is in a position to offer a unique perspective into initial assembly and growth. 

For our redshift range, overdensities are in general considered to be associated with the abundance of gas \citep[e.g.,][]{casey16, Noble17, Pensabene24}. There, galaxy star formation activity could be enhanced. The presence of $high$ \ewh\ emitters ($\simgt1000$\,\AA), not only of wELGs, in overdensities supports the idea. 
For galaxies to have a systematically reduced SFR in such gas abundance environments, {\it even stronger} regulation of star formation would be required. 

For the observed established continuum break seen in many wELGs, regulation by negative feedback from the recent starburst seems to be a straightforward consideration. In particular, considering the relatively small masses of the wELG samples ($\logm \sim 8.4$--9.5), they are more vulnerable to such feedback than typical quenched galaxies ($\logm\gg10$) are. What is not obvious is if supernova feedback from a single starburst could quench subsequent star formation activity. 
``Mini-quenching" is one of recently proposed scenarios \citep{dome24}, to explain the presence of similarly low-mass galaxies with a reduced star formation rate in the early universe \citep{looser23,strait23}. A toy model indicates that a single star formation event can halt star formation, but only lasts for $\sim20$--40\,Myr \citep[also][]{faisst24}. If overdensity environments somehow extend the duration of quenching, it would reasonably explain the observed high frequency of wELGs in the two overdensities, aside from any potential contribution from AGN, which we revisit in Sec.~\ref{sec:agn}.

On the other hand, since their masses are relatively low, it takes less environment heating to remove/heat up gas in an individual galaxy, quenching it. However, given the available timescale by the observed redshift, environmental quenching via, e.g., ram pressure stripping due to galaxies moving within the ICM and interactions after infall \citep[e.g.,][]{peng10,wetzel12} is not expected to be the dominant quenching channel. In fact, none of our galaxies show clear evidence of ram-pressure stripping in the NIRCam images. In a massive halo, the gas-heating mechanisms of virial shock can prevent the falling gas from reaching the center \citep[e.g.,][]{dekel06,daddi22}. However, the estimated halo masses of the two overdensities could be too small for this to happen, either. The two overdensities are located in the predicted regime, where cold gas can still penetrate through hot media and reach the central galaxy (Figure~\ref{fig:Mhalo}).

It is worth noting that studies of higher redshift (proto-) clusters have shown consistent pictures but somewhat at a weaker significance. \citet{morishita23b} reported that the member galaxies of A2744ODz7p9, a protocluster at $z=7.88$, have recent, short-duration star formation of an order of $\Delta t \simlt100$\,Myr. \citet{hashimoto23} reported that the member galaxies of the same protocluster have a systematically redder UV slope $\beta_{\rm UV}$ than field samples. {\citet{helton24} found a positive correlation between galaxy overdensity and $\beta_{\rm UV}$ using a spectroscopic sample of galaxies at $z=4.9$--$8.9$.} \citet[][]{li24} confirmed a similar trend using a large sample of galaxies in various environments over a wider redshift range. However, given that the $\beta_{\rm UV}$ measurement is sensitive to the enhancement of dust, it was unclear whether the redder slope can be attributed to the reduction of star formation in those member galaxies. In fact, most of the A2744ODz7p9 members show disturbed morphologies, indicative of recent/ongoing mergers. Direct measurements of \ha\ and continuum break strength in this study, on the other hand, have offered a more straightforward insight into the decline in star formation in overdensities. \citet[][]{witten24} recently reported a strong Balmer break and weak emission lines in one of the protocluster member galaxies, similar to the population of wELGs reported in our study.

\begin{figure}
\centering
    \includegraphics[width=0.48\textwidth]{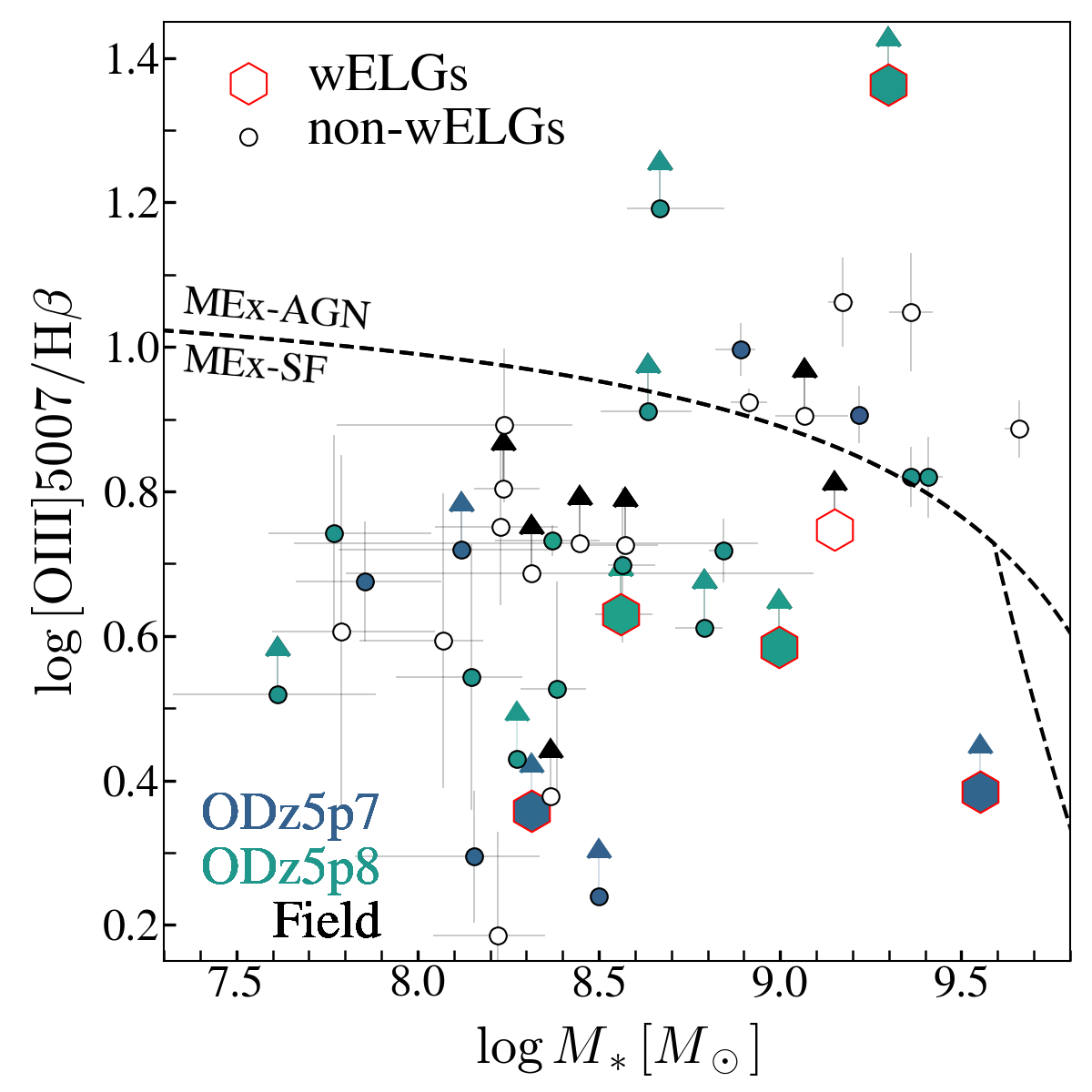}
	\caption{
 Our samples in the stellar mass-excitation (MEx) diagram. For those with \hb\ undetected, a 2-$\sigma$ lower limit is shown. Colors and symbols are as shown in Figure~\ref{fig:ew}. We do not see the dominance of potential AGN in the overdensity samples.
 }
\label{fig:mex}
\end{figure}

\subsection{The Potential Contributions by AGN on the Reduced Star Formation Activities}\label{sec:agn}

The presence of AGN is critical to understanding the mechanism responsible for the reduction in star formation activity inferred from \ewh. 
However, with the spectral resolution of the PRISM spectra, our ability to identify AGN candidates is limited. To further investigate the presence of AGN, we show our samples on the Mass-Excitation (MEx) diagram in Figure~\ref{fig:mex}, which offers an empirical classification for AGN \citep{juneau14}. 

Firstly, among all wELGs, there is one that shows a significantly high line ratio (ID53709, \oiii/\hb\,$>25$ at $2\,\sigma$), which makes it a very likely candidate for AGN. In addition, the observed \ha+\nii\,/\hb\ ratio of the same source, $\simgt38$ at $2\,\sigma$, is extremely high, potentially driven by strong \nii\ emission. However, despite the high significance of emission line detection (Figure~\ref{fig:spec_wel}), the spectrum does not show evidence of line broadening. The continuum spectrum of this source is well characterized by the underlying evolved stellar populations. These characteristics are seen in typical Seyfert~2 galaxies. For the observed \oiii/\hb\ ratio, \nii/\ha\ is expected to be $\sim1$--3, by taking the measurements for the Sloan Seyfert~2 galaxies \citep[e.g.,][]{kauffmann03, malkan17}. This indicates that the actual star formation activity in the source is even smaller.

{The observed high \ha+\nii\,/\hb\ ratio would require significant attenuation, $A_V>6$\,mag, even assuming a high \nii/\ha\ ratio, 3. Interestingly, the dust attenuation inferred from the SED analysis is rather small ($A_V=0.1$\,mag). One possible explanation for those observed properties is the existence of dusty AGN or star-forming region, hosted by a galaxy of relatively dust-free and old stellar populations (i.e. responsible for the observed \De). The nature of this source, and other potentially high \oiii\,/\hb\ sources, is further investigated in a separate study (Takahashi, in prep.)}.

Secondly, we revisit those with strong Balmer break but with a higher \ewh\ than 100\,\AA\ (IDs 53527, 59993, 48996), including the one showing a broad line (ID53527). These sources are an important population for our understanding of the potential origin of wELGs. All the three objects above are classified as MEx-AGN. In fact, while they are not selected as wELGs, all three sources have relatively weak \ha\ emissions, with \ewh\ ranging $\sim200$--300\,\AA. 
Also noted is that AGN typically have a higher \nii/\ha\ ratio. As such, their actual star formation activity may be even more reduced than inferred from \ewh. 
Therefore, we speculate that these non-wELGs are likely intermediate populations, under the influence of AGN on their star formation activity, and may end up to more quiescent populations like wELGs. 

{However, the MEx diagram is calibrated only at lower redshifts, leaving the application of the same classification boundary to low-metallicity, high-ionizing galaxies to be validated \citep[e.g.,][]{matthee23}. Because of this and also because a large fraction of our samples have lower limits, we caution that their nature should be concluded further with additional diagnostics.}

We note that the connection between wELGs and non-wELGs appears to be what is seen in the local universe, especially for those with high \oiii-to-\hb\ ratio. In the Sloan sample, the right wing in the BPT diagram (i.e. high \oiii/\hb\ and \nii/\ha) is populated by both bona-fide AGN and `retired' galaxies \citep[e.g.,][]{cidfernandes10}. The mass range of the wELG samples is considered relatively smaller than the typical mass at which AGN become active and start influencing star formation over galactic scale \citep{dimatteo05,hopkins06,somerville08}. 
{Because of the relatively shallow potential wells, those galaxies are susceptible to stellar feedback, as well as environmental effects through interactions with other galaxies.}

On the other hand, numerical studies \citep[e.g.,][]{koudmani21} found that even at this low mass range AGN can be influential. Recent findings of potential overmassive BHs at this cosmic time also support this \citep{izumi21,kocevski23,ubler23}. Given the prevalence of AGN in low-mass galaxies found by recent JWST observations \citep{harikane23c,matthee23,maiolino23}, AGN could be an equally possible quenching scenario. 

A missing link is, then, if the occurrence of AGN is related to the nature of overdensities, which is under much debate \citep[e.g.,][]{overzier16,macuga19,gatica24}. Theoretical studies predict a higher AGN fraction in overdensities compared to average fields, due to increased gas fractions and higher merger rates {\citep[e.g.,][]{dimatteo12,costa14}}. {Observational evidence supports this scenario: several studies have reported enhanced AGN activity in high-redshift ($1<z<3$) protoclusters relative to the field (e.g., $z=1.6$, \citealt{Krishnan17}; $z=2.2$, \citealt{Tozzi22}; $z=3.1$, \citealt{Monson23}).} However, it remains unknown whether such distinctions appear as early as $t_{\rm cosmo}<1$\,Gyr, especially considering the presence of negative feedback \citep[e.g.,][]{stiavelli05hst,kim09}. Indeed, we observed one potential AGN (ID43506, i.e. Little Red Dot) in isolation. A robust determination of the AGN fraction in the early overdensities with a larger sample would provide an invaluable benchmark and help us understand their interplay with overdensities.

\section{Summary and Future Prospect}\label{sec:sum}

In this study, we identified two overdensities of galaxies at $z\sim5.7$ in the sightline of the galaxy cluster Abell~2744. Combined with the sensitive JWST spectroscopic data, these unique early overdensites allowed us to conduct an in-depth investigation of both the overdensities themselves and the individual member galaxies. 
Remarkably, we identified six galaxies within the overdensities that show weak \ha+\nii\ emissions ($25\pm7\,\%$). The fraction of such galaxies was found significantly lower ($6\pm2\,\%$) in field samples of the equivalent redshift range.
The large scatter observed in \ewh\ of the overdensity samples suggests that overdensities are places where star formation can be enhanced too.
Taken together, these observations suggest that environments may be influential to galaxy evolution as early as within the first billion years after the Big Bang, leading to the accelerated maturation of galaxies as seen in lower redshift clusters and groups. 
While the exact physical mechanisms at play have yet to be investigated, our observations support the idea that feedback from intense star formation and/or AGN self-regulates subsequent activities in overdensities, leading to an increased fraction of wELGs compared to the average field. While AGN may contribute to the observed quiescence, their dominance in overdense environments {over other cluster-specific mechanisms such as galaxy-galaxy interactions}, remains unclear.

The evolutionary link of the wELGs identified here to massive, quenched galaxy populations at lower redshifts, discovered by recent JWST spectroscopic studies \citep[][]{carnall23,degraaff24}, is another point of interest. Predicting the future evolution of individual systems is challenging. However, we note that the stellar mass of wELGs in our samples is small, and they are unlikely to be the direct progenitor of quenched populations of $\sim10^{11}\,M_\odot$ by $z\sim4$. In addition, given the low star formation rate of wELGs, the mass gap between the two populations seems challenging to fill solely with intrinsic star formation, unless there is an abundance of gas reservoirs available. An alternative scenario is that most of the member galaxies will merge into a single system, which would ease the tension. However, the timescale for this to happen is not obvious and needs verification through numerical calculations. Future searches for wELGs to a larger extent will give us an improved view of the origin of these systems. If we still fail to find more massive wELGs in a larger sample, it would suggest that the majority of massive quenched galaxy populations were either formed in isolated environments or through a number of mergers in early density peaks. Direct confirmation of this can also be achieved by investigating the surrounding environment of massive quenched galaxies at lower redshifts.

\input{cand_z58.tex}

\begin{figure*}
\centering
\includegraphics[width=0.48\textwidth]{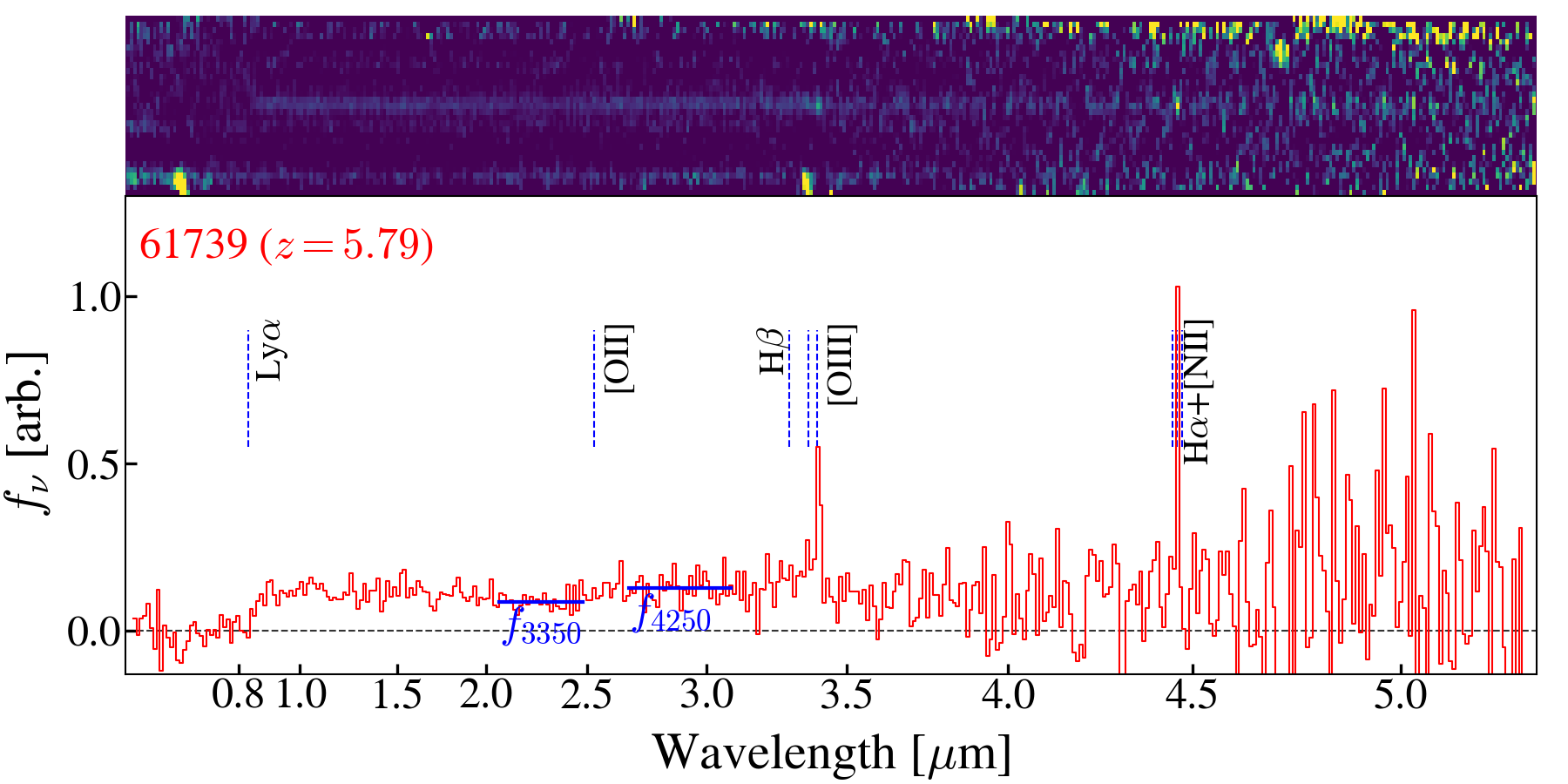}
\includegraphics[width=0.48\textwidth]{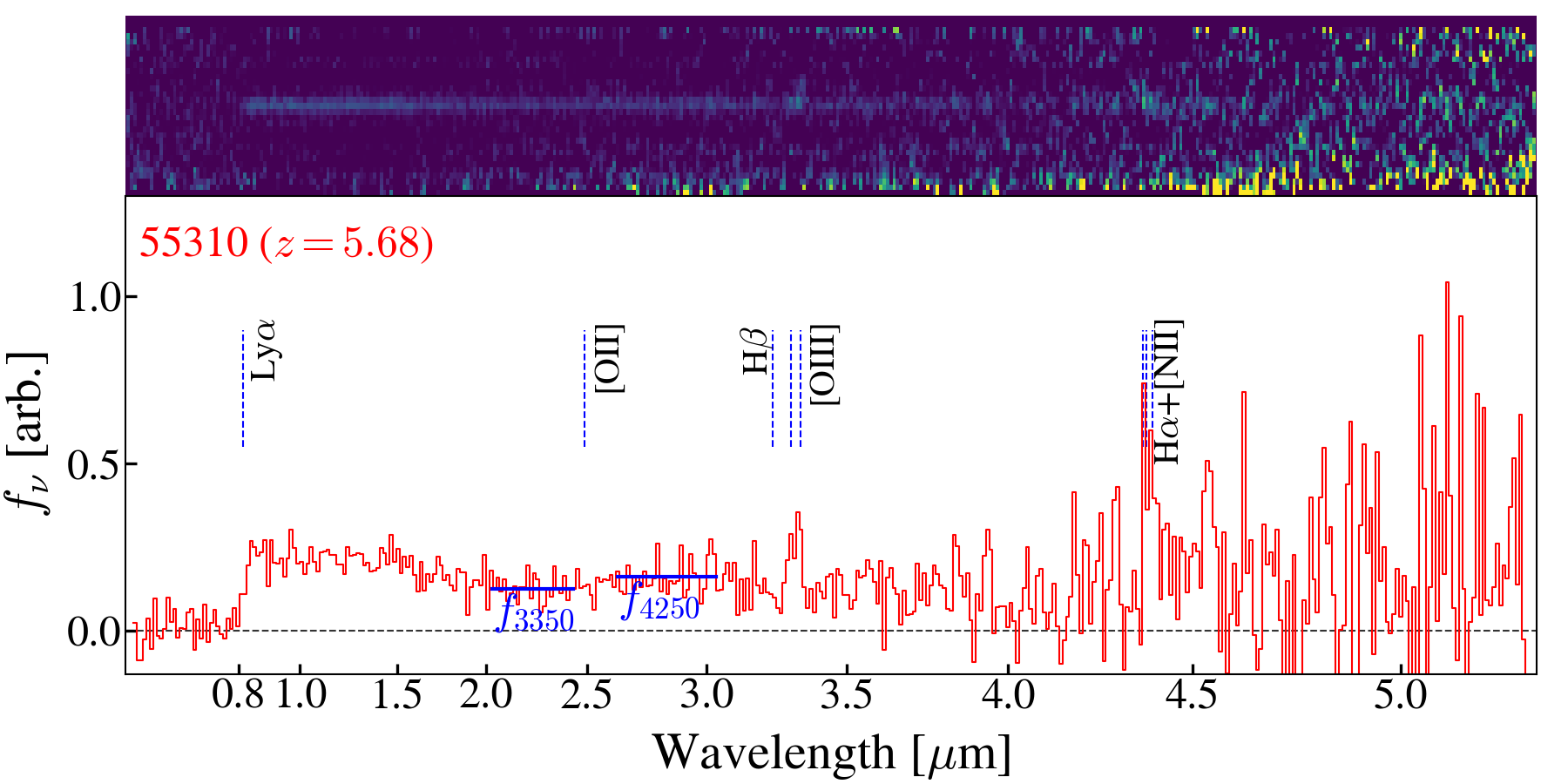}
\includegraphics[width=0.48\textwidth]{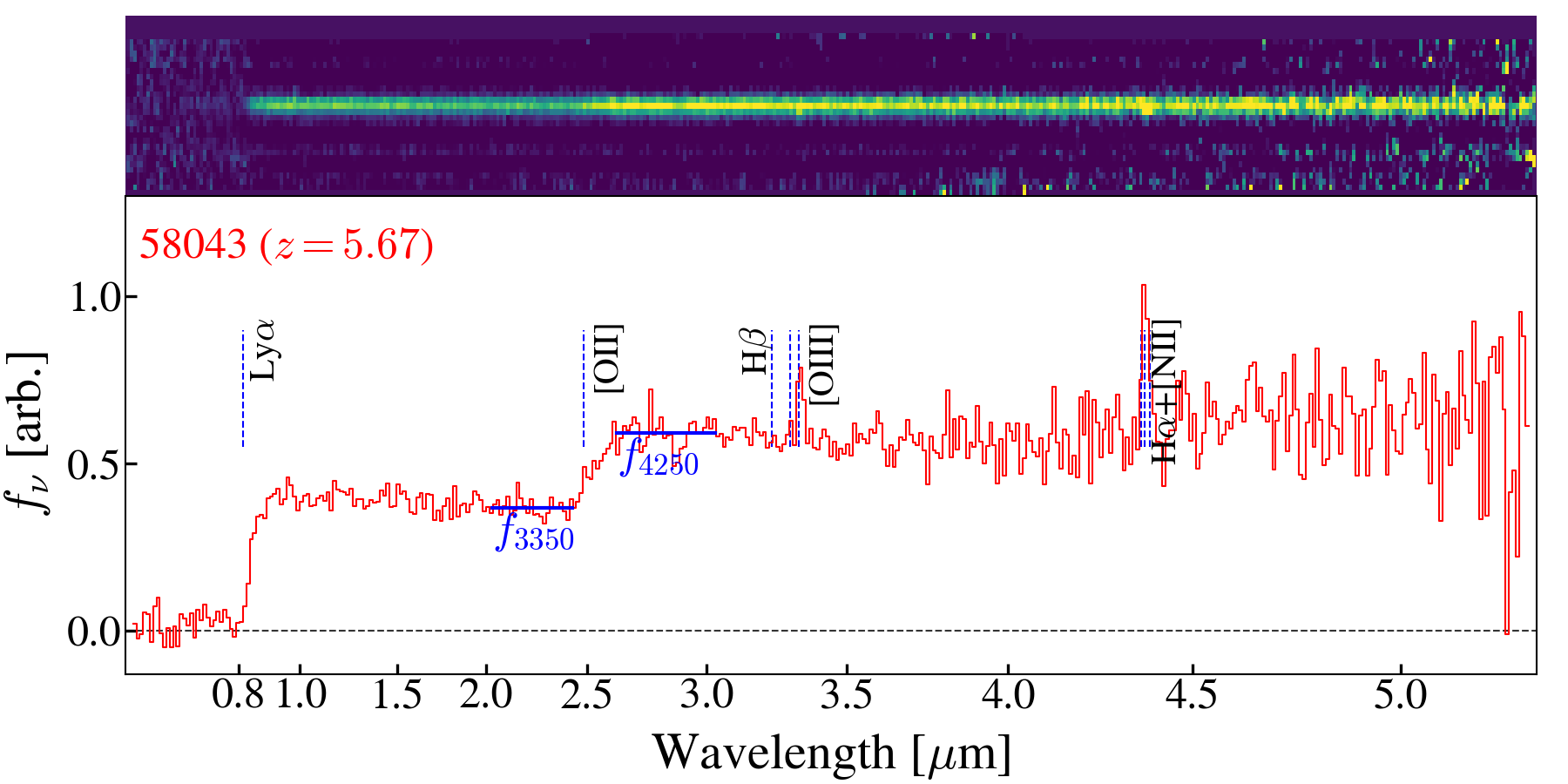}
\includegraphics[width=0.48\textwidth]{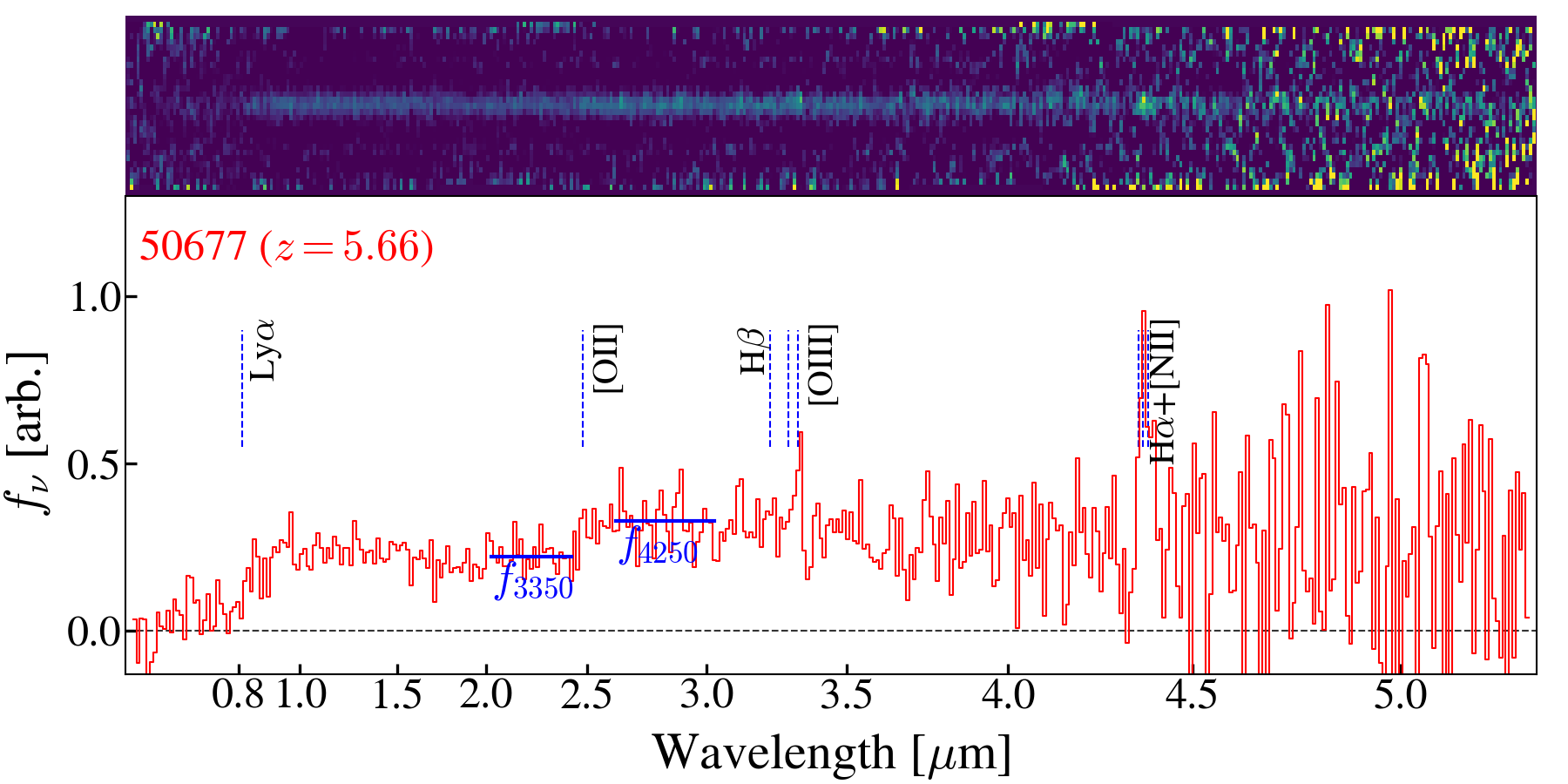}
\includegraphics[width=0.48\textwidth]{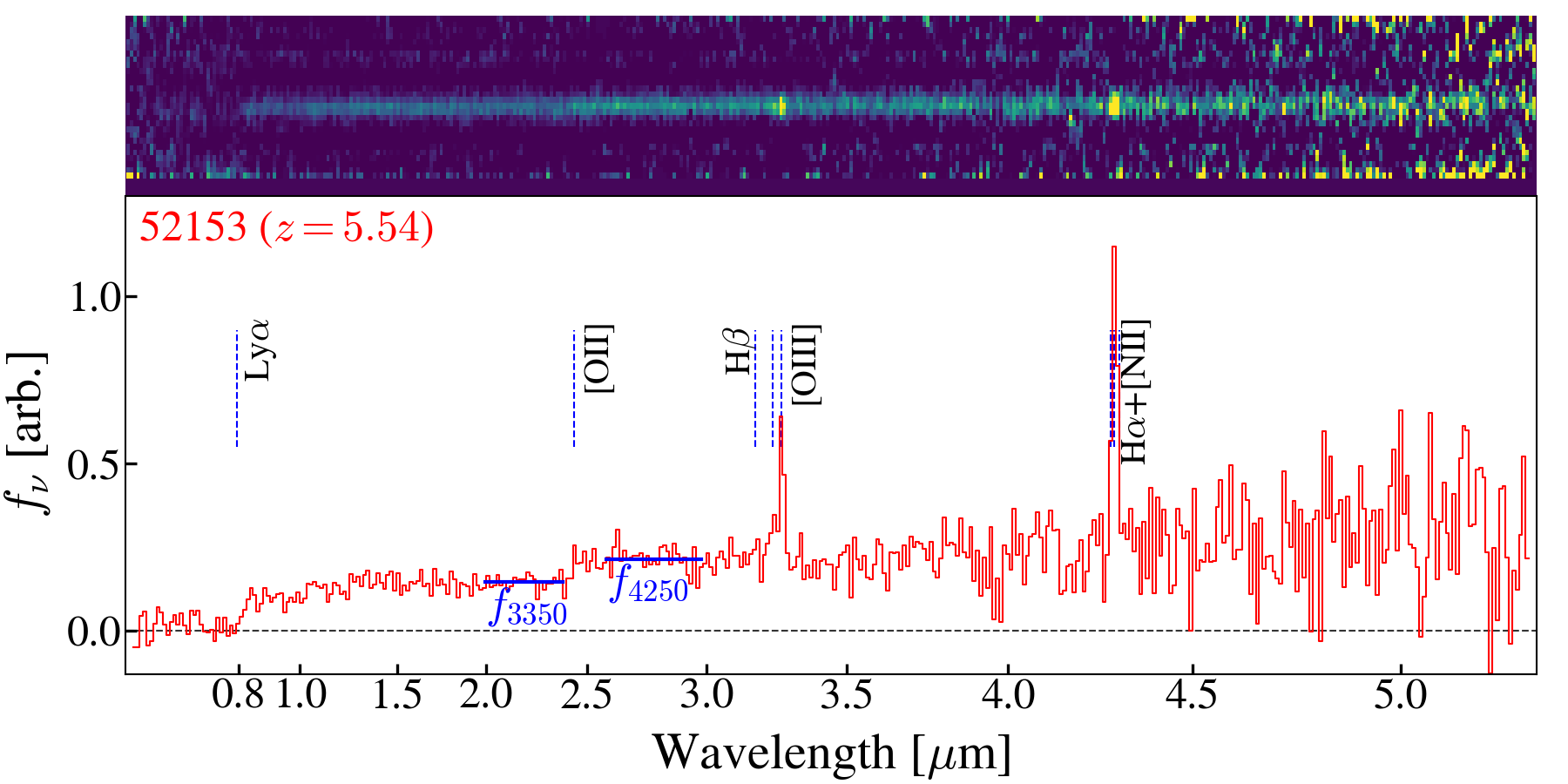}
\includegraphics[width=0.48\textwidth]{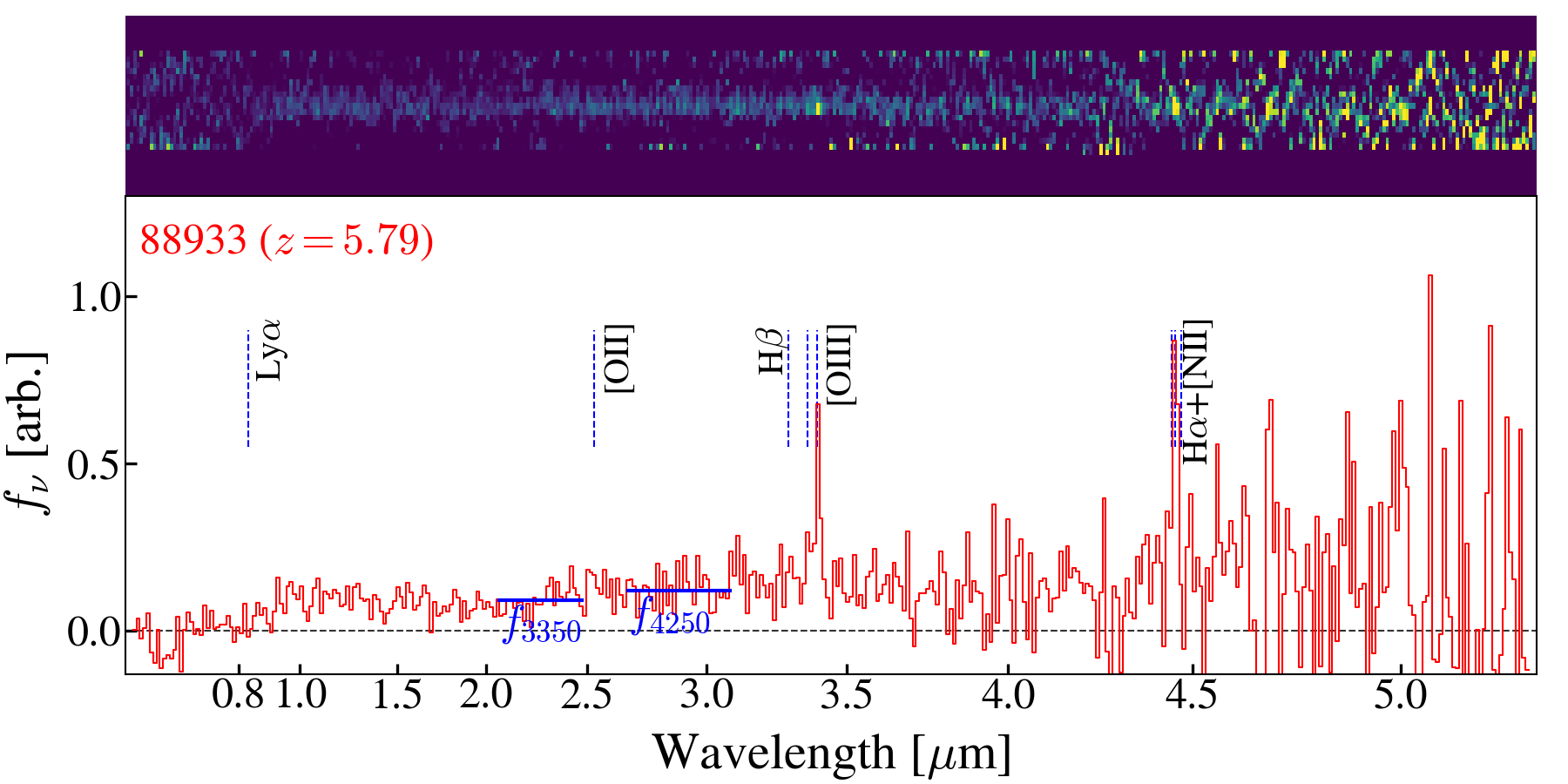}
 \caption{
Spectra of weak emission line galaxies (wELGs; Sec.~\ref{sec:ew}). The figure format is the same as in Figure~\ref{fig:spec_wel}.  
}
\label{fig:spec_wel_all}
\end{figure*}


\section*{Acknowledgements}
We thank the anonymous referee for their careful reading of the manuscript and constructive comments.
We acknowledge the teams of the JWST observation programs, IDs~1324, 2561, 2756, 2883, 3073, 3516, 3538, 3990, and 4110 for their dedicated work in designing and planning these programs and for generously making their data publicly available. We thank Seunghwan Lim for kindly sharing their halo mass trajectories from the FLAMINGO simulation. TM would like to thank Ranga-Ram Chary for his insightful comments on the manuscript. 
This work is based on observations made with the NASA/ESA/CSA {\it James Webb Space Telescope (JWST)}. The JWST data presented in this article were obtained from the Mikulski Archive for Space Telescopes (MAST) at the Space Telescope Science Institute. The specific observations analyzed can be accessed via \dataset[doi: 10.17909/j4rg-3f57]{https://doi.org/10.17909/j4rg-3f57} \citep{roberts-borsani24} and \dataset[doi: 10.17909/4r6b-bx96]{https://doi.org/10.17909/4r6b-bx96} (JWST-GO-3073). We acknowledge financial support by the PRIN 2022 MUR project 2022CB3PJ3 – First Light And Galaxy aSsembly (FLAGS) funded by the European Union – Next Generation EU, and by INAF Mini-grant ``Reionization and Fundamental Cosmology with High-Redshift Galaxies." This research is supported in part by the Australian Research Council Centre of Excellence for All Sky Astrophysics in 3 Dimensions (ASTRO 3D), through project number CE170100013. We also acknowledge support from the INAF Large Grant 2022 “Extragalactic Surveys with JWST” (PI Pentericci). B.V. is supported  by the European Union – NextGenerationEU RFF M4C2 1.1 PRIN 2022 project 2022ZSL4BL INSIGHT. Z.L. is supported by the KAKENHI Grant Number 24KJ0394 through the Japan Society for the Promotion of Science (JSPS).

{
{\it Software:} 
Astropy \citep{astropy13,astropy18,astropy22}, bbpn \citep{bbpn}, EAzY \citep{brammer08}, EMCEE \citep{foreman13}, gsf \citep{morishita19}, numpy \citep{numpy}, python-fsps \citep{foreman14}.
}


\bibliography{ms}{}
\bibliographystyle{aasjournal}



\end{document}

%% file: cand_z58.tex
\startlongtable
\begin{deluxetable*}{cccccccccccccc}
\tablecolumns{14}
\tabcolsep=0.06cm
\tablecaption{
}
\tablehead{
\colhead{ID} & \colhead{R.A.} & \colhead{Decl.} & \colhead{$z$} & \colhead{inst} & \colhead{texp} & \colhead{$\mu$} & \colhead{$\log M_*$} & \colhead{$\log {\rm SFR}$} & \colhead{$\log t$} & \colhead{$A_V$} & \colhead{EW$_0({\rm H\alpha{+}[\textrm{N}~\textsc{ii}]})$} & \colhead{$\log D_e 4000$} & \colhead{$\log$[\textrm{O}~\textsc{iii}]/$\mathrm{H\beta}$}\\
\colhead{} & \colhead{deg.} & \colhead{deg.} & \colhead{} & \colhead{} & \colhead{sec} & \colhead{} & \colhead{$M_\odot$} & \colhead{$M_\odot {\rm yr^{-1}}$} & \colhead{Gyr} & \colhead{mag} & \colhead{${\rm \AA}$} & \colhead{}
}
\startdata
\cutinhead{ODz5p7 ($5.640<z<5.680$)}
80595 & 3.51062 & -30.30537 & $5.6450$ & ${\rm msa}$ & $15755$ & $1.2_{-0.0}^{+0.0}$ & $9.2_{-0.0}^{+0.0}$ &  $0.3_{-0.0}^{+0.1}$ &  $-0.6_{-0.3}^{+0.2}$ &  $0.0_{-0.0}^{+0.0}$ & $>2760$ & $0.04\pm0.06$ & $0.9\pm0.0$\\
50956 & 3.60621 & -30.38662 & $5.6550$ & ${\rm wfss}$ & $47413$ & $2.0_{-0.0}^{+0.1}$ & $9.4_{-0.1}^{+0.0}$ &  $0.9_{-0.0}^{+0.1}$ &  $-0.9_{-0.1}^{+0.1}$ &  $0.1_{-0.0}^{+0.0}$ &  &  & \\
49662 & 3.62365 & -30.38427 & $5.6556$ & ${\rm wfss}$ & $29375$ & $1.5_{-0.0}^{+0.0}$ & $9.2_{-0.0}^{+0.0}$ &  $0.7_{-0.2}^{+0.1}$ &  $-1.0_{-0.1}^{+0.1}$ &  $0.2_{-0.1}^{+0.1}$ &  &  & \\
53131 & 3.59074 & -30.37940 & $5.6565$ & ${\rm wfss}$ & $44321$ & $2.8_{-0.1}^{+0.1}$ & $8.6_{-0.1}^{+0.1}$ &  $0.8_{-0.1}^{+0.1}$ &  $-1.5_{-0.2}^{+0.1}$ &  $0.1_{-0.1}^{+0.0}$ &  &  & \\
55048 & 3.60194 & -30.35852 & $5.6570$ & ${\rm msa}$ & $15755$ & $1.9_{-0.1}^{+0.1}$ & $8.5_{-0.1}^{+0.1}$ &  $0.4_{-0.8}^{+0.9}$ &  $-1.3_{-0.5}^{+0.4}$ &  $0.4_{-0.2}^{+0.3}$ & $>165$ & $0.34\pm0.05$ & --\\
53596 & 3.58709 & -30.37909 & $5.6570$ & ${\rm wfss}$ & $49475$ & $3.2_{-0.1}^{+0.1}$ & $7.8_{-0.3}^{+0.2}$ &  $0.3_{-0.2}^{+0.2}$ &  $-1.7_{-0.3}^{+0.3}$ &  $0.2_{-0.1}^{+0.1}$ &  &  & \\
49572 & 3.62461 & -30.38420 & $5.6573$ & ${\rm wfss}$ & $24222$ & $1.5_{-0.0}^{+0.0}$ & $9.0_{-0.1}^{+0.1}$ &  $0.7_{-0.2}^{+0.3}$ &  $-1.3_{-0.2}^{+0.2}$ &  $0.2_{-0.1}^{+0.1}$ &  &  & \\
56326 & 3.55999 & -30.38286 & $5.6590$ & ${\rm wfss}$ & $33498$ & $3.9_{-0.3}^{+0.3}$ & $8.6_{-0.1}^{+0.1}$ &  $0.7_{-0.1}^{+0.2}$ &  $-1.1_{-0.1}^{+0.1}$ &  $0.1_{-0.1}^{+0.1}$ &  &  & \\
50677 & 3.62013 & -30.37873 & $5.6590$ & ${\rm msa}$ & $15755$ & $1.6_{-0.0}^{+0.0}$ & $8.5_{-0.3}^{+0.2}$ &  $0.6_{-0.8}^{+0.7}$ &  $-1.5_{-0.4}^{+0.3}$ &  $0.7_{-0.3}^{+0.3}$ & $<92$$^\dagger$ & $0.17\pm0.03$ & --\\
51298 & 3.59069 & -30.39555 & $5.6600$ & ${\rm msa}$ & $78507$ & $10.0_{-0.6}^{+0.8}$ & $8.9_{-0.1}^{+0.0}$ &  $1.5_{-0.0}^{+0.0}$ &  $-1.2_{-0.1}^{+0.1}$ &  $0.0_{-0.0}^{+0.0}$ & $578\pm70$ & $0.01\pm0.03$ & $1.0\pm0.0$\\
51543 & 3.57611 & -30.40446 & $5.6600$ & ${\rm wfss}$ & $45352$ & $2.9_{-0.1}^{+0.1}$ & $8.6_{-0.2}^{+0.1}$ &  $0.8_{-0.0}^{+0.1}$ &  $-1.5_{-0.1}^{+0.1}$ &  $0.0_{-0.0}^{+0.0}$ &  &  & \\
58759 & 3.58166 & -30.35060 & $5.6650$ & ${\rm msa}$ & $15755$ & $4.4_{-0.6}^{+0.9}$ & $7.9_{-0.2}^{+0.2}$ &  $0.6_{-0.4}^{+0.4}$ &  $-1.8_{-0.3}^{+0.3}$ &  $0.3_{-0.1}^{+0.2}$ & $>862$ & $0.03\pm0.10$ & $0.7\pm0.1$\\
54365 & 3.60026 & -30.36428 & $5.6670$ & ${\rm msa}$ & $15755$ & $2.0_{-0.1}^{+0.1}$ & $8.2_{-0.3}^{+0.2}$ &  $0.1_{-0.3}^{+0.3}$ &  $-1.4_{-0.5}^{+0.5}$ &  $0.2_{-0.1}^{+0.1}$ & $>5731$ & $-0.00\pm0.08$ & $0.3\pm0.1$\\
56103 & 3.59928 & -30.35404 & $5.6670$ & ${\rm msa}$ & $15755$ & $2.1_{-0.1}^{+0.1}$ & $8.1_{-0.3}^{+0.2}$ &  $0.4_{-0.3}^{+0.3}$ &  $-1.7_{-0.4}^{+0.4}$ &  $0.3_{-0.1}^{+0.1}$ & $>1650$ & $0.10\pm0.15$ & $>0.7$\\
55321 & 3.59436 & -30.36236 & $5.6672$ & ${\rm wfss}$ & $35044$ & $2.4_{-0.1}^{+0.2}$ & $8.5_{-0.1}^{+0.1}$ &  $0.5_{-0.7}^{+0.8}$ &  $-1.3_{-0.3}^{+0.2}$ &  $0.5_{-0.2}^{+0.3}$ &  &  & \\
58043 & 3.59316 & -30.34647 & $5.6680$ & ${\rm msa}$ & $15755$ & $2.0_{-0.1}^{+0.1}$ & $9.6_{-0.0}^{+0.0}$ &  $1.3_{-0.1}^{+0.1}$ &  $-1.3_{-0.3}^{+0.2}$ &  $0.1_{-0.0}^{+0.0}$ & $30\pm6$$^\dagger$ & $0.20\pm0.01$ & --\\
55310 & 3.59431 & -30.36251 & $5.6760$ & ${\rm msa}$ & $52862$ & $2.4_{-0.1}^{+0.2}$ & $8.3_{-0.1}^{+0.1}$ &  $0.5_{-0.4}^{+0.5}$ &  $-1.6_{-0.3}^{+0.3}$ &  $0.3_{-0.1}^{+0.2}$ & $<73$$^\dagger$ & $0.11\pm0.04$ & --\\
\cutinhead{ODz5p8 ($5.764<z<5.804$)}
50110 & 3.57303 & -30.41955 & $5.7642$ & ${\rm wfss}$ & $24737$ & $1.8_{-0.0}^{+0.0}$ & $7.9_{-0.1}^{+0.1}$ &  $0.2_{-0.2}^{+0.2}$ &  $-1.8_{-0.3}^{+0.4}$ &  $0.1_{-0.0}^{+0.1}$ &  &  & \\
50863 & 3.57063 & -30.41463 & $5.7650$ & ${\rm msa}$ & $75871$ & $1.8_{-0.0}^{+0.0}$ & $9.4_{-0.0}^{+0.0}$ &  $1.2_{-0.1}^{+0.1}$ &  $-1.4_{-0.3}^{+0.2}$ &  $0.2_{-0.1}^{+0.1}$ & $130\pm20$ & -- & $0.8\pm0.0$\\
53429 & 3.56483 & -30.39717 & $5.7694$ & ${\rm wfss}$ & $22160$ & $2.2_{-0.1}^{+0.1}$ & $8.0_{-0.1}^{+0.2}$ &  $0.4_{-0.2}^{+0.2}$ &  $-1.6_{-0.2}^{+0.2}$ &  $0.1_{-0.1}^{+0.1}$ &  &  & \\
49882 & 3.60542 & -30.39658 & $5.7710$ & ${\rm msa}$ & $62488$ & $2.2_{-0.1}^{+0.1}$ & $7.8_{-0.2}^{+0.3}$ &  $0.1_{-0.1}^{+0.2}$ &  $-1.7_{-0.2}^{+0.2}$ &  $0.1_{-0.1}^{+0.1}$ & $>1176$ & $-0.30\pm0.45$ & $0.7\pm0.1$\\
89694 & 3.47607 & -30.31035 & $5.7710$ & ${\rm msa}$ & $15755$ & $1.2_{-0.0}^{+0.0}$ & $8.1_{-0.2}^{+0.1}$ &  $-0.2_{-0.7}^{+0.7}$ &  $-1.4_{-0.4}^{+0.5}$ &  $0.3_{-0.2}^{+0.2}$ & $>606$ & $0.11\pm0.09$ & $0.5\pm0.2$\\
46821 & 3.58522 & -30.43137 & $5.7710$ & ${\rm msa}$ & $15755$ & $1.7_{-0.0}^{+0.0}$ & $7.6_{-0.3}^{+0.3}$ &  $0.2_{-0.4}^{+0.4}$ &  $-2.2_{-0.2}^{+0.4}$ &  $0.2_{-0.1}^{+0.1}$ & $>586$ & $0.17\pm0.10$ & $>0.5$\\
53423 & 3.56492 & -30.39715 & $5.7717$ & ${\rm wfss}$ & $19583$ & $2.2_{-0.1}^{+0.1}$ & $8.6_{-0.1}^{+0.1}$ &  $0.8_{-0.1}^{+0.1}$ &  $-1.6_{-0.2}^{+0.2}$ &  $0.1_{-0.1}^{+0.1}$ &  &  & \\
48996 & 3.61314 & -30.39705 & $5.7720$ & ${\rm msa}$ & $15755$ & $2.0_{-0.0}^{+0.0}$ & $8.7_{-0.1}^{+0.2}$ &  $0.6_{-0.1}^{+0.1}$ &  $-1.4_{-0.3}^{+0.2}$ &  $0.1_{-0.0}^{+0.1}$ & $>310$ & $0.18\pm0.10$ & $>1.2$\\
53846 & 3.56520 & -30.39427 & $5.7730$ & ${\rm msa}$ & $51048$ & $2.4_{-0.1}^{+0.1}$ & $8.8_{-0.0}^{+0.0}$ &  $1.4_{-0.2}^{+0.1}$ &  $-1.9_{-0.3}^{+0.2}$ &  $0.3_{-0.1}^{+0.0}$ & $935\pm32$ & -- & $0.7\pm0.0$\\
59993 & 3.55409 & -30.36362 & $5.7740$ & ${\rm msa}$ & $87048$ & $2.7_{-0.2}^{+0.2}$ & $8.6_{-0.1}^{+0.1}$ &  $0.5_{-0.4}^{+0.5}$ &  $-1.2_{-0.4}^{+0.4}$ &  $0.3_{-0.1}^{+0.3}$ & $243\pm54$ & $0.22\pm0.03$ & $>0.9$\\
93770 & 3.45971 & -30.31249 & $5.7750$ & ${\rm msa}$ & $15755$ & $1.2_{-0.0}^{+0.0}$ & $8.3_{-0.0}^{+0.0}$ &  $-0.2_{-0.1}^{+0.2}$ &  $-1.1_{-0.3}^{+0.2}$ &  $0.1_{-0.0}^{+0.1}$ & $<157$ & $0.07\pm0.06$ & --\\
48689 & 3.61923 & -30.39446 & $5.7750$ & ${\rm wfss}$ & $24737$ & $1.7_{-0.0}^{+0.0}$ & $9.6_{-0.0}^{+0.0}$ &  $1.3_{-0.1}^{+0.2}$ &  $-1.3_{-0.3}^{+0.2}$ &  $0.2_{-0.1}^{+0.1}$ &  &  & \\
50756 & 3.61025 & -30.38559 & $5.7756$ & ${\rm wfss}$ & $47928$ & $1.8_{-0.0}^{+0.1}$ & $9.0_{-0.0}^{+0.0}$ &  $1.0_{-0.1}^{+0.1}$ &  $-1.6_{-0.2}^{+0.2}$ &  $0.1_{-0.0}^{+0.0}$ &  &  & \\
84406 & 3.48715 & -30.31302 & $5.7790$ & ${\rm msa}$ & $15755$ & $1.2_{-0.0}^{+0.0}$ & $8.8_{-0.1}^{+0.1}$ &  $-0.1_{-0.3}^{+0.3}$ &  $-0.8_{-0.3}^{+0.2}$ &  $0.2_{-0.1}^{+0.1}$ & $>396$ & -- & $>0.6$\\
53527 & 3.56109 & -30.39926 & $5.7790$ & ${\rm msa}$ & $15755$ & $1.9_{-0.1}^{+0.1}$ & $9.4_{-0.1}^{+0.0}$ &  $1.3_{-0.1}^{+0.1}$ &  $-1.3_{-0.2}^{+0.1}$ &  $0.4_{-0.1}^{+0.1}$ & $236\pm27$ & $0.26\pm0.01$ & $0.8\pm0.1$\\
53486 & 3.56110 & -30.39932 & $5.7799$ & ${\rm wfss}$ & $28860$ & $1.9_{-0.1}^{+0.1}$ & $9.1_{-0.0}^{+0.0}$ &  $1.2_{-0.1}^{+0.1}$ &  $-1.5_{-0.2}^{+0.1}$ &  $0.1_{-0.0}^{+0.0}$ &  &  & \\
94381 & 3.46566 & -30.30868 & $5.7830$ & ${\rm msa}$ & $15755$ & $1.2_{-0.0}^{+0.0}$ & $8.6_{-0.0}^{+0.1}$ &  $0.3_{-0.1}^{+0.1}$ &  $-1.5_{-0.3}^{+0.2}$ &  $0.1_{-0.1}^{+0.0}$ & $>891$ & $-0.02\pm0.07$ & $0.7\pm0.1$\\
53709 & 3.59614 & -30.37138 & $5.7840$ & ${\rm msa}$ & $15755$ & $2.3_{-0.1}^{+0.1}$ & $9.3_{-0.0}^{+0.0}$ &  $0.6_{-0.1}^{+0.1}$ &  $-0.7_{-0.3}^{+0.2}$ &  $0.1_{-0.1}^{+0.1}$ & $107\pm7$$^\dagger$ & $0.28\pm0.01$ & $>1.4$\\
93880 & 3.46578 & -30.30868 & $5.7850$ & ${\rm msa}$ & $15755$ & $1.2_{-0.0}^{+0.0}$ & $8.4_{-0.1}^{+0.1}$ &  $0.3_{-0.1}^{+0.1}$ &  $-1.7_{-0.2}^{+0.2}$ &  $0.2_{-0.1}^{+0.1}$ & $>1051$ & $0.27\pm0.12$ & $0.5\pm0.1$\\
88933 & 3.46682 & -30.31909 & $5.7900$ & ${\rm msa}$ & $15755$ & $1.2_{-0.0}^{+0.0}$ & $9.0_{-0.1}^{+0.0}$ &  $0.7_{-0.1}^{+0.1}$ &  $-1.5_{-0.2}^{+0.2}$ &  $0.2_{-0.1}^{+0.1}$ & $89\pm38$$^\dagger$ & $0.12\pm0.06$ & $>0.6$\\
51797 & 3.56117 & -30.41380 & $5.7920$ & ${\rm wfss}$ & $48959$ & $1.6_{-0.0}^{+0.0}$ & $8.3_{-0.3}^{+0.3}$ &  $0.3_{-0.8}^{+0.9}$ &  $-1.6_{-0.3}^{+0.3}$ &  $0.4_{-0.1}^{+0.5}$ &  &  & \\
61739 & 3.57488 & -30.33753 & $5.7950$ & ${\rm msa}$ & $15755$ & $1.9_{-0.1}^{+0.1}$ & $8.6_{-0.1}^{+0.1}$ &  $0.7_{-0.5}^{+0.5}$ &  $-1.7_{-0.3}^{+0.4}$ &  $0.5_{-0.2}^{+0.3}$ & $73\pm25$$^\dagger$ & $0.16\pm0.05$ & $>0.6$\\
58700 & 3.54281 & -30.38065 & $5.7950$ & ${\rm msa}$ & $15755$ & $2.7_{-0.2}^{+0.3}$ & $8.4_{-0.2}^{+0.1}$ &  $1.2_{-0.1}^{+0.1}$ &  $-2.1_{-0.2}^{+0.2}$ &  $0.4_{-0.0}^{+0.0}$ & $1553\pm230$ & $0.07\pm0.05$ & $0.7\pm0.0$\\
53055 & 3.55982 & -30.40371 & $5.7958$ & ${\rm wfss}$ & $28860$ & $1.8_{-0.0}^{+0.0}$ & $8.0_{-0.3}^{+0.2}$ &  $-0.2_{-0.5}^{+0.5}$ &  $-1.4_{-0.5}^{+0.5}$ &  $0.2_{-0.1}^{+0.2}$ &  &  & \\
51150 & 3.59144 & -30.39667 & $5.8034$ & ${\rm wfss}$ & $49475$ & $10.6_{-0.6}^{+0.8}$ & $7.7_{-0.1}^{+0.1}$ &  $0.4_{-0.1}^{+0.2}$ &  $-1.5_{-0.3}^{+0.2}$ &  $0.0_{-0.0}^{+0.1}$ &  &  & \\
\cutinhead{Field Sample}
68601 & 3.51938 & -30.35446 & $5.5250$ & ${\rm msa}$ & $15755$ & $1.8_{-0.1}^{+0.1}$ & $8.1_{-0.2}^{+0.1}$ &  $0.1_{-0.3}^{+0.4}$ &  $-1.5_{-0.3}^{+0.3}$ &  $0.2_{-0.1}^{+0.2}$ & $290\pm129$ & $-0.25\pm0.09$ & $0.6\pm0.2$\\
71184 & 3.50563 & -30.34941 & $5.5250$ & ${\rm msa}$ & $15755$ & $1.5_{-0.0}^{+0.0}$ & $8.2_{-0.2}^{+0.2}$ &  $0.3_{-0.4}^{+0.4}$ &  $-1.7_{-0.2}^{+0.3}$ &  $0.5_{-0.1}^{+0.2}$ & $>833$ & $0.04\pm0.08$ & $0.8\pm0.1$\\
52836 & 3.61310 & -30.36461 & $5.5269$ & ${\rm wfss}$ & $7215$ & $1.6_{-0.0}^{+0.1}$ & $9.0_{-0.0}^{+0.0}$ &  $1.0_{-0.1}^{+0.1}$ &  $-1.5_{-0.1}^{+0.1}$ &  $0.2_{-0.0}^{+0.0}$ &  &  & \\
55293 & 3.60640 & -30.35339 & $5.5350$ & ${\rm msa}$ & $15755$ & $1.7_{-0.1}^{+0.1}$ & $8.4_{-0.6}^{+0.2}$ &  $0.3_{-1.3}^{+1.3}$ &  $-1.6_{-0.5}^{+0.5}$ &  $0.9_{-0.4}^{+0.6}$ & $>75$ & -- & --\\
49775 & 3.58499 & -30.41307 & $5.5360$ & ${\rm wfss}$ & $24737$ & $3.5_{-0.1}^{+0.1}$ & $8.2_{-0.1}^{+0.1}$ &  $0.3_{-0.1}^{+0.1}$ &  $-1.1_{-0.1}^{+0.1}$ &  $0.1_{-0.0}^{+0.0}$ &  &  & \\
52153 & 3.57006 & -30.40369 & $5.5450$ & ${\rm msa}$ & $55616$ & $2.1_{-0.0}^{+0.1}$ & $9.2_{-0.0}^{+0.0}$ &  $1.2_{-0.1}^{+0.1}$ &  $-1.5_{-0.2}^{+0.1}$ &  $0.4_{-0.0}^{+0.0}$ & $92\pm16$$^\dagger$ & $0.17\pm0.02$ & $>0.7$\\
51129 & 3.56063 & -30.42030 & $5.5513$ & ${\rm wfss}$ & $36590$ & $1.5_{-0.0}^{+0.0}$ & $9.2_{-0.1}^{+0.0}$ &  $0.6_{-0.1}^{+0.1}$ &  $-0.8_{-0.2}^{+0.1}$ &  $0.1_{-0.0}^{+0.0}$ &  &  & \\
51377 & 3.56250 & -30.41654 & $5.5516$ & ${\rm wfss}$ & $48444$ & $1.6_{-0.0}^{+0.0}$ & $8.4_{-0.1}^{+0.1}$ &  $0.0_{-0.2}^{+0.3}$ &  $-1.1_{-0.2}^{+0.2}$ &  $0.1_{-0.1}^{+0.1}$ &  &  & \\
79252 & 3.50683 & -30.31160 & $5.5880$ & ${\rm msa}$ & $15755$ & $1.3_{-0.0}^{+0.0}$ & $8.2_{-0.2}^{+0.1}$ &  $0.5_{-0.6}^{+0.4}$ &  $-2.1_{-0.4}^{+0.4}$ &  $0.5_{-0.2}^{+0.1}$ & $>257$ & $0.41\pm0.55$ & $0.2\pm0.1$\\
88370 & 3.46098 & -30.32513 & $5.5930$ & ${\rm msa}$ & $15755$ & $1.2_{-0.0}^{+0.0}$ & $8.9_{-0.1}^{+0.1}$ &  $1.0_{-0.0}^{+0.0}$ &  $-1.7_{-0.1}^{+0.1}$ &  $0.1_{-0.0}^{+0.0}$ & $767\pm102$ & $-0.01\pm0.04$ & $0.9\pm0.0$\\
70459 & 3.51550 & -30.34624 & $5.5990$ & ${\rm msa}$ & $15755$ & $1.5_{-0.0}^{+0.0}$ & $9.4_{-0.1}^{+0.1}$ &  $1.1_{-0.2}^{+0.2}$ &  $-1.3_{-0.1}^{+0.1}$ &  $0.5_{-0.1}^{+0.1}$ & $520\pm97$ & $0.10\pm0.03$ & $1.0\pm0.1$\\
48753 & 3.63058 & -30.38553 & $5.6080$ & ${\rm msa}$ & $15755$ & $1.5_{-0.0}^{+0.0}$ & $8.3_{-0.5}^{+0.8}$ &  $0.3_{-3.4}^{+3.2}$ &  $-1.6_{-0.8}^{+1.1}$ &  $1.2_{-0.8}^{+1.7}$ & $>104$ & $-0.07\pm0.43$ & $>0.7$\\
57147 & 3.57595 & -30.36545 & $5.6123$ & ${\rm wfss}$ & $25768$ & $5.0_{-0.5}^{+0.6}$ & $8.6_{-0.1}^{+0.1}$ &  $0.6_{-0.1}^{+0.1}$ &  $-0.9_{-0.2}^{+0.1}$ &  $0.1_{-0.0}^{+0.0}$ &  &  & \\
60258 & 3.55926 & -30.35749 & $5.6204$ & ${\rm wfss}$ & $49990$ & $2.5_{-0.2}^{+0.2}$ & $9.3_{-0.0}^{+0.0}$ &  $1.4_{-0.2}^{+0.2}$ &  $-1.5_{-0.2}^{+0.2}$ &  $0.3_{-0.1}^{+0.1}$ &  &  & \\
60089 & 3.56818 & -30.35228 & $5.6230$ & ${\rm msa}$ & $63071$ & $3.2_{-0.3}^{+0.6}$ & $7.8_{-0.2}^{+0.3}$ &  $0.6_{-0.9}^{+0.8}$ &  $-2.0_{-0.5}^{+0.6}$ &  $0.7_{-0.3}^{+0.4}$ & $355\pm162$ & $-0.06\pm0.08$ & $0.6\pm0.2$\\
57351 & 3.57583 & -30.36532 & $5.6243$ & ${\rm wfss}$ & $24222$ & $5.3_{-0.6}^{+0.7}$ & $8.2_{-0.1}^{+0.1}$ &  $0.7_{-0.1}^{+0.1}$ &  $-1.5_{-0.1}^{+0.1}$ &  $0.1_{-0.0}^{+0.0}$ &  &  & \\
51117 & 3.62066 & -30.37426 & $5.6270$ & ${\rm msa}$ & $19878$ & $1.6_{-0.0}^{+0.0}$ & $9.7_{-0.0}^{+0.0}$ &  $1.4_{-0.1}^{+0.1}$ &  $-1.3_{-0.1}^{+0.1}$ &  $0.2_{-0.0}^{+0.0}$ & $303\pm26$ & $0.11\pm0.01$ & $0.9\pm0.0$\\
60298 & 3.58610 & -30.33732 & $5.6320$ & ${\rm msa}$ & $15755$ & $1.8_{-0.1}^{+0.1}$ & $8.4_{-0.8}^{+0.5}$ &  $0.2_{-1.8}^{+2.0}$ &  $-1.6_{-0.7}^{+0.8}$ &  $0.6_{-0.4}^{+1.2}$ & $>171$ & $0.03\pm0.12$ & $>0.7$\\
50110 & 3.57303 & -30.41955 & $5.7642$ & ${\rm wfss}$ & $24737$ & $1.8_{-0.0}^{+0.0}$ & $7.9_{-0.1}^{+0.1}$ &  $0.2_{-0.2}^{+0.2}$ &  $-1.8_{-0.3}^{+0.4}$ &  $0.1_{-0.0}^{+0.1}$ &  &  & \\
43506 & 3.61920 & -30.42326 & $5.8370$ & ${\rm msa}$ & $15755$ & $1.6_{-0.0}^{+0.0}$ & $6.9_{-0.3}^{+0.1}$ &  $0.1_{-0.2}^{+0.1}$ &  $-2.5_{-0.3}^{+0.2}$ &  $0.1_{-0.0}^{+0.0}$ & $432\pm24$ & $0.35\pm0.05$ & $0.4\pm0.0$\\
53139 & 3.61317 & -30.36218 & $5.8420$ & ${\rm msa}$ & $15755$ & $1.6_{-0.0}^{+0.1}$ & $8.2_{-0.5}^{+0.2}$ &  $-0.1_{-0.6}^{+0.7}$ &  $-1.6_{-0.6}^{+0.6}$ &  $0.2_{-0.2}^{+0.3}$ & $>1548$ & $-0.07\pm0.08$ & $0.9\pm0.1$\\
69718 & 3.51380 & -30.35333 & $5.9250$ & ${\rm msa}$ & $15755$ & $1.7_{-0.0}^{+0.1}$ & $8.2_{-0.1}^{+0.1}$ &  $0.5_{-0.2}^{+0.2}$ &  $-1.9_{-0.2}^{+0.2}$ &  $0.3_{-0.1}^{+0.1}$ & $>496$ & $0.01\pm0.09$ & $>0.8$\\
67911 & 3.55368 & -30.33006 & $5.9290$ & ${\rm msa}$ & $15755$ & $1.5_{-0.0}^{+0.0}$ & $9.1_{-0.1}^{+0.1}$ &  $0.6_{-0.2}^{+0.3}$ &  $-1.0_{-0.2}^{+0.2}$ &  $0.2_{-0.1}^{+0.2}$ & $244\pm73$ & $0.19\pm0.04$ & $>0.9$\\
69619 & 3.51813 & -30.35045 & $5.9320$ & ${\rm msa}$ & $15755$ & $1.7_{-0.1}^{+0.1}$ & $9.2_{-0.0}^{+0.0}$ &  $1.0_{-0.2}^{+0.1}$ &  $-1.4_{-0.3}^{+0.2}$ &  $0.2_{-0.1}^{+0.1}$ & $495\pm66$ & $0.07\pm0.02$ & $1.1\pm0.1$\\
60260 & 3.55733 & -30.35957 & $5.9340$ & ${\rm wfss}$ & $57205$ & $2.5_{-0.1}^{+0.2}$ & $8.3_{-0.1}^{+0.1}$ &  $0.3_{-0.3}^{+0.4}$ &  $-1.5_{-0.3}^{+0.3}$ &  $0.2_{-0.1}^{+0.1}$ &  &  & \\
55866 & 3.57494 & -30.37400 & $5.9347$ & ${\rm wfss}$ & $41229$ & $3.4_{-0.2}^{+0.2}$ & $8.6_{-0.0}^{+0.0}$ &  $0.5_{-0.3}^{+0.2}$ &  $-1.3_{-0.3}^{+0.2}$ &  $0.4_{-0.1}^{+0.1}$ &  &  & \\
72143 & 3.49244 & -30.35283 & $5.9650$ & ${\rm msa}$ & $15755$ & $1.4_{-0.0}^{+0.0}$ & $8.6_{-0.1}^{+0.0}$ &  $0.3_{-0.2}^{+0.2}$ &  $-1.2_{-0.3}^{+0.2}$ &  $0.1_{-0.1}^{+0.1}$ & $>475$ & -- & --\\
55777 & 3.56770 & -30.37909 & $5.9920$ & ${\rm wfss}$ & $47928$ & $3.6_{-0.2}^{+0.2}$ & $9.3_{-0.0}^{+0.0}$ &  $1.1_{-0.1}^{+0.1}$ &  $-0.7_{-0.3}^{+0.2}$ &  $0.1_{-0.0}^{+0.0}$ &  &  & \\
78798 & 3.49097 & -30.32492 & $5.9970$ & ${\rm msa}$ & $15755$ & $1.3_{-0.0}^{+0.0}$ & $8.6_{-0.1}^{+0.1}$ &  $0.1_{-0.4}^{+0.4}$ &  $-1.1_{-0.3}^{+0.3}$ &  $0.2_{-0.1}^{+0.2}$ & $>341$ & $-0.14\pm0.24$ & $>0.7$\\
\enddata
\tablecomments{
Measurements are corrected for magnification (Sec.~\ref{sec:data}).
$\dagger$: weak emission line galaxies, wELGs (Sec.~\ref{sec:ew}). 
$\ddagger$: Little Red Dot. SED parameters represent the UV-bright component.
Among MSA samples, those with \hb\ and \oiii\ undetected are shown with ``--''. 
84406 and 72143 do not have the wavelength coverage for measuring $D_{e}4000$. 
55293 is not detected in continuum and thus $D_{e}4000$ is not available. 50863 and 53846 are covered by the high-resolution grating and therefore $D_{e}4000$ is not available.
}\label{tab:phys}
\end{deluxetable*}